\documentclass[aps,prl,amsmath,twocolumn,superscriptaddress,floatfix]{revtex4-1}
\usepackage{amsmath}
\usepackage{amssymb}
\usepackage{graphicx}
\usepackage{bbold}
\usepackage{float}
\usepackage[colorlinks=true,urlcolor=blue,anchorcolor=blue,linkcolor=blue,citecolor=blue,breaklinks=true]{hyperref}
\usepackage{adjustbox}
\usepackage{ulem}
\makeatletter
\usepackage{booktabs,multirow}
\usepackage{stmaryrd}
\usepackage[table]{xcolor}
\makeatother

\definecolor{applegreen}{rgb}{0.55, 0.71, 0.0}
\definecolor{magneta}{rgb}{1.0,0.0,1.0}
\definecolor{nb}{rgb}{0.0,0.550,0.85}

\newcommand\skout{\bgroup\markoverwith{\textcolor{red}{\rule[0.5ex]{2pt}{0.9pt}}}\ULon}

\definecolor{applegreen}{rgb}{0.55, 0.71, 0.0}

\begin{document}

\title{Electronic conduction and superconducting properties of CoSi$_2$ films on silicon\\ -- an unconventional superconductor with technological potential}

\author{Shao-Pin Chiu}
\email{Present address: Department of Physics, Fu Jen Catholic University, Taipei 24205, Taiwan}
\affiliation{Department of Electrophysics, National Yang-  Ming Chiao Tung University, Hsinchu 30010, Taiwan}

\author{Chang-Jan Wang}
\affiliation{Institute of Physics, National Yang Ming Chiao Tung University, Hsinchu 30010, Taiwan}

\author{Yi-Chun Lin}
\affiliation{Department of Electrophysics, National Yang Ming Chiao Tung University, Hsinchu 30010, Taiwan}

\author{Shun-Tast Tu}
\affiliation{Institute of Physics, National Yang Ming Chiao Tung University, Hsinchu 30010, Taiwan}

\author{Shouray Sahu}
\affiliation{International College of Semiconductor Technology, National Yang Ming Chiao Tung University, Hsinchu 30010, Taiwan}

\author{Ruey-Tay Wang}
\affiliation{Department of Electrophysics, National Yang Ming Chiao Tung University, Hsinchu 30010, Taiwan}

\author{Chih-Yuan Wu}
\affiliation{Department of Physics, Fu Jen Catholic University, Taipei 24205, Taiwan}

\author{Sheng-Shiuan Yeh}
\affiliation{International College of Semiconductor Technology, National Yang Ming Chiao Tung University, Hsinchu 30010, Taiwan}

\author{Stefan Kirchner}
\affiliation{Department of Electrophysics, National Yang Ming Chiao Tung University, Hsinchu 30010, Taiwan}

\author{Juhn-Jong Lin}
\email{Corresponding author. E-mail address: jjlin@nycu.edu.tw (J. J. Lin).}
\affiliation{Department of Electrophysics, National Yang Ming Chiao Tung University, Hsinchu 30010, Taiwan}

\date{\today}
\begin{abstract}
We report observations of unusual normal-state electronic conduction properties and superconducting characteristics of high-quality CoSi$_2$/Si films grown on silicon Si(100) and Si(111) substrates. A good understanding of these features shall help to address the underlying physics of the unconventional pairing symmetry recently observed in transparent CoSi$_2$/TiSi$_2$ heterojunctions [S. P. Chiu \textit{et al.}, Sci. Adv. \textbf{7}, eabg6569 (2021); Nanoscale \textbf{15}, 9179 (2023)], where CoSi$_2$/Si is a superconductor with a superconducting transition temperature $T_c \simeq$ (1.1--1.5) K, dependent on its dimensions, and TiSi$_2$ is a normal metal. In CoSi$_2$/Si films, we find a pronounced positive magnetoresistance caused by the weak-antilocalization effect, indicating a strong Rashba spin-orbit coupling (SOC). This SOC generates two-component  superconductivity in CoSi$_2$/TiSi$_2$ heterojunctions. The CoSi$_2$/Si films are stable under ambient conditions and have ultralow 1/$f$ noise. Moreover, they can be patterned via the standard lithography techniques, which might be of considerable practical value for future scalable superconducting and quantum device fabrication.\\ 


\end{abstract}

\maketitle

    
\section{Introduction}
	
The physics of unconventional superconducting states has been front and center of recent theoretical and experimental condensed matter research \cite{Qi2011,Kallin2016,Stewart2017,Linder2019}. Both scientific interest and potential technological prospects drive this development. Recently, through phase-sensitive conductance spectroscopy studies of the ``anomalous proximity effect" \cite{Tanaka2004anomalous,Tanaka2005theory}, we have found that the Cooper pairing characteristics in nonmagnetic CoSi$_2$/TiSi$_2$ superconductor/normal metal heterojunctions point to spin-triplet pairing symmetry \cite{Chiu.21,Chiu2023}, where CoSi$_2$ is a superconductor (S) with an optimal superconducting transition temperature $T_c \approx$ 1.55 K, and TiSi$_2$ is a normal metal (N) down to at least 50 mK. More precisely, in the vicinity of the high-transmittance CoSi$_2$/TiSi$_2$ S/N interface, the pair function has a dominant even-frequency spin-triplet odd- parity $p$-wave component in CoSi$_2$/Si. In contrast, it has an odd-frequency spin-triplet even-parity $s$-wave component in TiSi$_2$. Our S/N heterojunctions were grown on silicon Si(100) or Si(111) substrates. The device fabrication method has been described elsewhere \cite{Chiu2021b} and is briefly summarized in Section II. Because the unusual physical properties arise from the thin-film form of CoSi$_2$ on silicon, we shall denote our films by CoSi$_2$/Si in this paper. We emphasize that the unconventional nature of the superconductivity observed in Refs. \cite{Chiu.21} and \cite{Chiu2023} is only pertinent to S/N junctions made of CoSi$_2$/Si films. Bulk (poly- or single-crystalline) CoSi$_2$ is a conventional BCS $s$-wave superconductor. The pairing in bulk CoSi$_2$ is mediated by the electron-phonon interaction \cite{tsutsumi1995}, with an electron-phonon coupling constant $\lambda \simeq$ 0.44 and a renormalized Coulomb pseudopotential between electrons $\mu^\ast \simeq$ 0.13 \cite{Allen1993}.

The conductance spectra of two-terminal S/N junctions [a schematic is provided in Fig. \ref{fig1}(a)] and three-terminal so-called T-shaped superconducting proximity structures\cite{Asano2007}, schematically depicted in Fig. \ref{fig1}(c), are both phase sensitive to the pairing symmetry of the S component. These two kinds of hetero-devices can provide independent and complementary information to discriminate an unconventional superconductor from a conventional one. In the T-shaped proximity structure, the S is attached through an N arm and located a short distance (on the order of the superconducting coherence length) away from the normal-metal wire N. Extensive theoretical calculations have been conducted by Tanaka et al. \cite{Tanaka2007} and Asano et al. \cite{Asano2007} who concluded that  T-shaped superconducting proximity structures act as diagnostic tools for $p$-wave pairing. The main panel of Fig. \ref{fig1}(a) depicts that for a S/N junction, the differential conductance, defined by $G(V,T) = dI(V,T)/dV$, features a familiar zero-bias conductance dip for a spin-singlet $s$-wave S component, where $I$ is the current, $V$ is the bias voltage, and $T$ is the temperature. In sharp contrast, the theory predicts a broad zero-bias conductance peak (ZBCP) accompanied by two symmetric side dips for a spin-triplet $p$-wave pairing S component. The two side dips signify the superconducting energy gap, $\triangle (T)$. Furthermore, for the T-shaped proximity structure where $G(V)$ curves are measured along the N wire [cf. Fig. \ref{fig1}(c)], the theory predicts a zero-bias conductance dip for a spin-singlet S component, while it predicts a ZBCP for a spin-triplet S component. The width of the ZBCP in this case is small, being on the order of the Thouless energy $E_{\rm Th}$ [$\ll \triangle_0 = \triangle (T$\,$\rightarrow$\,0)]. Consequently, conductance spectroscopy studies can be a powerful probe for distinguishing the $p$-wave from conventional $s$-wave pairing symmetry in the S component.

\begin{widetext}
\begin{center}
\begin{figure}[t!]
\centering
\includegraphics[width= 0.8\textwidth]{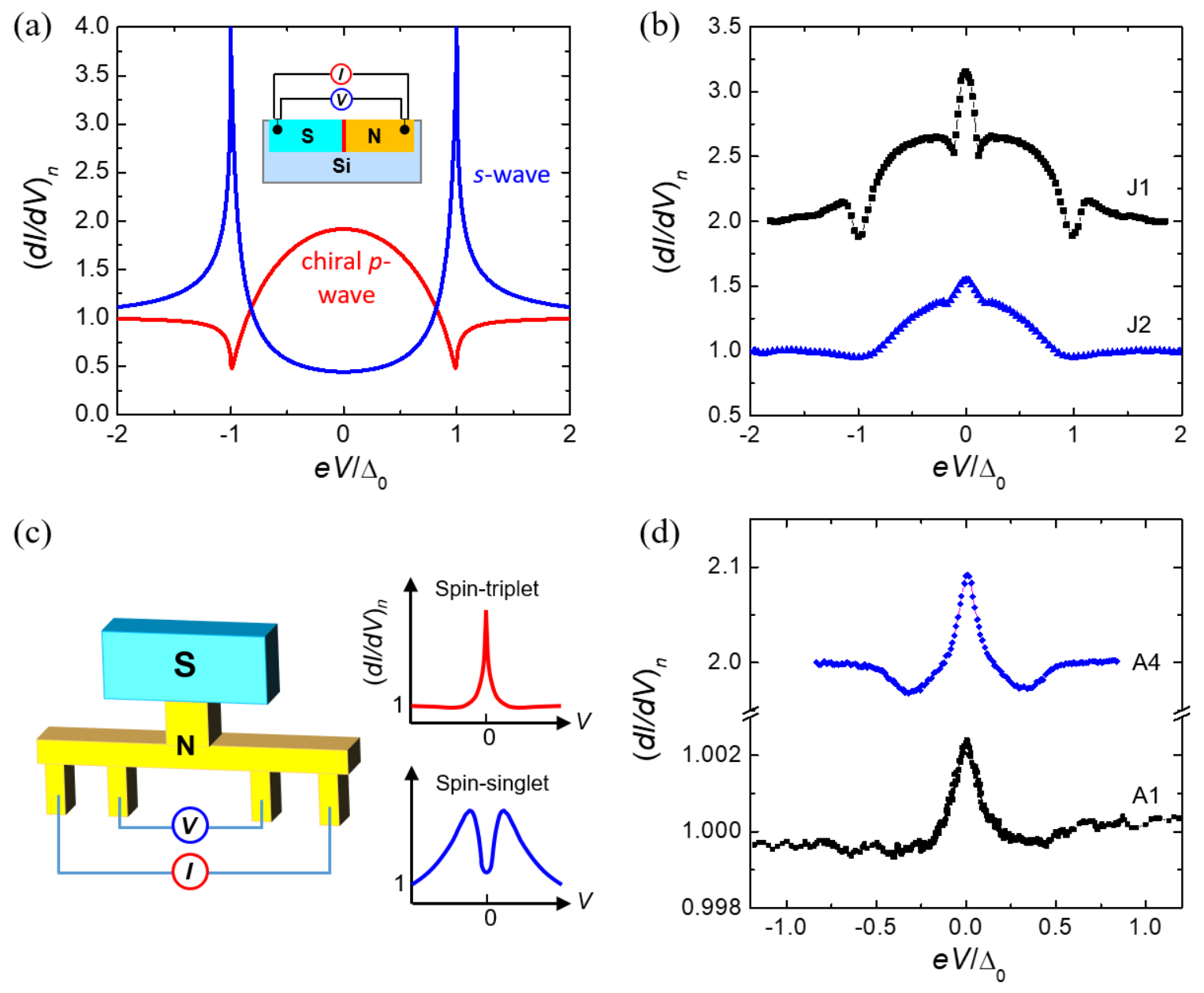} 
\caption{Conductance spectra of S/N heterojunctions and T-shaped superconducting proximity structures. (a) Schematic normalized conductance spectra $(dI/dV)_n$ for a S/N junction of barrier strength $\tilde{Z}$ = 1 with a $s$-wave (blue) and chiral $p$-wave (red) pairing S component \cite{BTK1982,Yamashiro1997}. Inset: A schematic CoSi$_2$/TiSi$_2$ S/N heterojunction on a silicon substrate with a 4-probe configuration. (b) $(dI/dV)_n$ versus $eV/\triangle_0$ for two representative CoSi$_2$/TiSi$_2$ S/N heterojunctions manifesting $p$-wave pairing (taken from \cite{Chiu.21}). (c) Left: A schematic T-shaped superconducting proximity structure with a 4-probe configuration. Right: Schematic conductance spectra of a T-shaped proximity device with a spin-triplet (top) and spin-singlet (bottom) S component. (d) $(dI/dV)_n$ versus $eV/\triangle_0$ for two CoSi$_2$/TiSi$_2$ T-shaped proximity devices manifesting spin-triplet pairing (device A1 taken from \cite{Chiu.21}). In (b) and (d), the $(dI/dV)_n$ curves were measured at $T$ = 0.37 K and in zero magnetic field. They are vertically offset for clarity.
}
\label{fig1}
\end{figure}
\end{center}
\end{widetext}

Experimentally, it is challenging to fabricate S/N junctions and T-shaped proximity devices with a clean S/N interface so that the Andreev reflection can occur \cite{Courtois1999}. Figures \ref{fig1}(b) and \ref{fig1}(d) respectively show the normalized $(dI/dV)_n$ curves versus $eV/\triangle_0$ for two CoSi$_2$/TiSi$_2$ S/N junctions and two T-shaped proximity devices which we have recently measured \cite{Chiu.21}, where $(dI/dV)_n$ is the $G(V)$ normalized to its corresponding normal-state value, defined by $(dI/dV)_n = [dI(V,T)/dV]/[dI(V,T$=4\,K)$/dV]$, and $e$ is the electronic charge. In both cases, the line shape of our conductance spectra conforms to the spin-triplet pairing symmetry. This is a very encouraging observation because the spin-triplet pairing symmetry is nontrivial to realize in real materials \cite{Kallin2016,Stewart2017}. Among elemental metals, metal alloys and compound superconductors, besides He$^3$~\cite{VollhardtWoefle}, triplet pairing has only been found in a limited number of materials, including UTe$_2$, URhGe, UPt$_3$ and UBe$_{13}$ \cite{Stewart2017}, while two-component superconductivity exists in MnSi, some heavy fermion compounds, and other non-inversion symmetric superconductors \cite{Shimizu2019,Schemm.14,Frigeri2004,Bauer2004}. Two-component nematic superconductivity in 4Hb-TaS$_2$ has been reported recently \cite{Silber.24}. In Fig. \ref{fig1}(b), the central narrow peak on top of the broad hump is an extraordinary feature compatible with expectations for chiral $p$-wave pairing, which becomes more pronounced when a thin regime in the N component adjacent to the S/N interface has a high resistance relative to the junction resistance \cite{Tanaka2005theory}. Apart from the conductance spectra, our measurements of the magnetoresistance (MR) of small S/N junctions manifest unusual ``advanced" hysteresis in low magnetic fields (not shown), a phenomenon likely arising from net spontaneous supercurrents associated with chiral $p$-wave domains \cite{Chiu.21}. This unusual signature provides further evidence for an unconventional superconducting state in CoSi$_2$/TiSi$_2$ heterojunctions.

This paper reports the normal-state electronic conduction properties and superconducting characteristics for a series of CoSi$_2$/Si films with thickness ($t$) ranging from 5 to 315 nm. A thorough understanding of this intriguing thin-film material's fundamental transport features may help unravel the microscopic origin(s) for the unconventional superconductivity found in CoSi$_2$/TiSi$_2$ heterojunctions. In addition, because our junction fabrication processes are fully compatible with the present-date silicon-based integrated circuit (IC) technology, micro-fabrication and scalability of CoSi$_2$/Si-based superconducting circuits should be feasible. Thus, it is desirable to study the $t$ dependence of the transport and superconducting properties to understand which $t$ range would be most relevant and suitable for potential quantum-technology applications. For comparison and completeness, we have also fabricated several TiSi$_2$/Si films and several NiSi$_2$/Si films and studied their transport characteristics. These two disilicides are normal metals. They have much weaker spin-orbit coupling (SOC) than that in CoSi$_2$/Si films. 

This paper is organized as follows. Section II contains our experimental methods for sample fabrication and electrical transport measurements. Section III includes the electronic conduction properties, including measurements of spin-orbit scattering time and the electron dephasing time. The low-frequency 1/$f$ noise in the normal state is presented. The transport properties of TiSi$_2$/Si films and NiSi$_2$/Si films are also studied for comparison. Section IV discusses the variation of superconducting properties with $t$, and estimates the superconducting coherence length and penetration depth. Our conclusion is given in section V.

\section{Experimental method}
\subsection{Growth of Disilicide films}

\textbf{Growth of CoSi$_2$/Si films.}
A thin Co layer of thickness $t_{\rm Co}$ was deposited on a high-purity Si(100) substrate or a Si(111) substrate via thermal evaporation. The 99.998\% purity Co wire was supplied by Alfa Aesar Corporation. Intrinsic (\textit{i.e.}, nominally undoped) Si wafers with $\rho$(300\,K)\,$>$\,20000 $\Omega$ cm were made by the floating zone method and supplied by Summit-tech corporation. A mechanical mask was used to define the sample geometry. In some cases, the electron-beam lithography and lift-off technique was utilized to make small patterns. The deposited Co on silicon was thermally annealed at several hundred degrees $^\circ$C and for a few hours in a high vacuum ($\approx$\,3\,$\times$\,$10^{-6}$ Torr) to achieve the desired CoSi$_2$ phase. The final thickness of the single-phased CoSi$_2$/Si film was previously empirically established to be $t \simeq 3.5\, t_{\rm Co}$ \cite{Ommen1988,Chen2004}. The optimal annealing temperature ($T_A$) to form the stoichiometric CoSi$_2$ phase with minimal agglomeration ranged from 550$^\circ$C to 800$^\circ$C. The thinner the as-deposited $t_{\rm Co}$ was, the lower the optimal $T_A$ value preferred. The duration time for annealing at the optimal $T_A$ temperature typically ranged from 0.5 to 2 h.

\begin{figure}[t!]
\begin{center}
\includegraphics[width=0.48\textwidth]{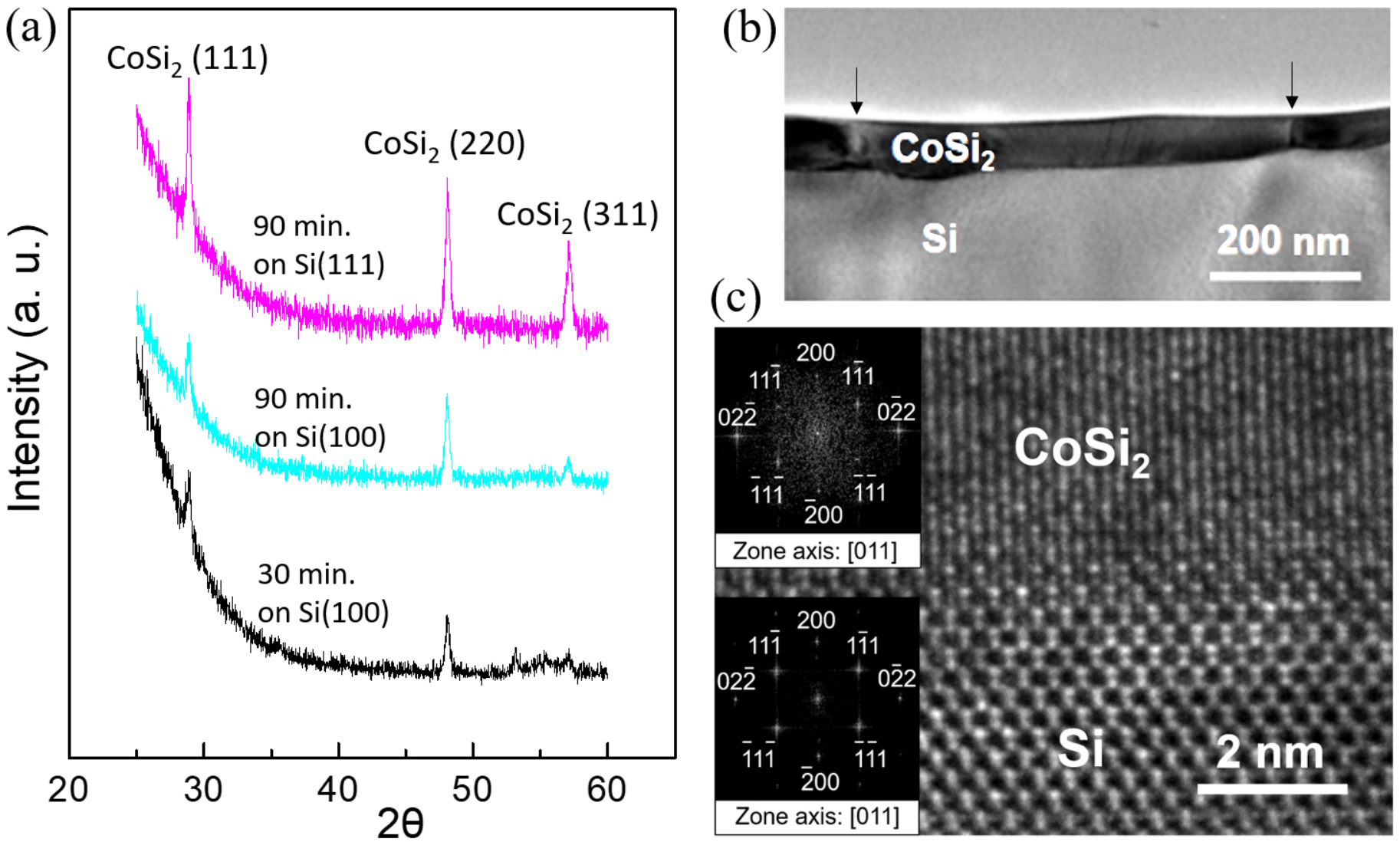} 
\caption{Structure characterizations of CoSi$_2$/Si films annealed at 800$^\circ$C.
(a) GIXRD spectra for three as-grown 105-nm thick CoSi$_2$/Si films. The films were annealed for 30 or 90 min. (b) Cross-sectional TEM image for a 90-min annealed CoSi$_2$/Si(100) film. The arrows indicate grain boundaries. (c) Zoom-in image of atomically sharp CoSi$_2$/Si(100) interface. Left top and bottom insets: TEM diffraction patterns for CoSi$_2$ and Si(100), respectively. They reveal the epitaxial relation of CoSi$_2$(100) on Si(100) from the information: CoSi$_2$[011]\,$\sslash$\,Si[011] and CoSi$_2$(200)\,$\sslash$\,Si(200).
}
\label{fig2}
\end{center}
\end{figure}

Due to the minor lattice mismatch ($\approx$\,$-$1.2\%) of CoSi$_2$ to Si, epitaxial phases of CoSi$_2$ on the Si(100) or Si(111) substrate were readily formed under proper thermal annealing conditions. The grazing incidence X-ray diffraction (GIXRD) spectra for three as-grown 105-nm thick CoSi$_2$/Si films are plotted in Fig \ref{fig2}(a). The spectra reveal the preferred lattice orientations of (111), (220), and (311), confirming the formation of the CoSi$_2$ phase on silicon. The 30-min annealed film has a barely visible (311) peak, reflecting an insufficient annealing time period. 

Figure \ref{fig2}(b) shows the cross-sectional transmission electron microscopy (TEM) image for a $\approx$\,100-nm thick CoSi$_2$ film on Si(100). The two arrows indicate the grain boundaries of a CoSi$_2$ grain with a lateral size of $\approx$\,400 nm. Figure \ref{fig2}(c) shows a high-resolution TEM image demonstrating the epitaxy of CoSi$_2$(100) on Si(100). It clearly reveals that CoSi$_2$ and Si form a sharp and strain-free metal-semiconductor interface. Two kinds of epitaxial relations often compete in the CoSi$_2$/Si(100) system, namely, the epitaxy of CoSi$_2$(100) on Si(100) and the epitaxy of CoSi$_2$(110) on Si(100), as previously established \cite{Chen1991,Bulle1992} and discussed elsewhere \cite{Chiu.17}. 

\textbf{Growth of TiSi$_2$/Si and NiSi$_2$/Si films.}
Using similar growth processes, we have also fabricated a series of TiSi$_2$/Si films and a series of NiSi$_2$/Si films and measured their electrical transport properties. In particular, we have utilized TiSi$_2$ for the \textit{in situ} fabrication of CoSi$_2$/TiSi$_2$ heterojunctions to achieve high S/N interface transmittance, which is required for studying the conductance spectra arising from the Andreev reflection \cite{BTK1982}. We find that TiSi$_2$/Si films have resistivity values similar to those of CoSi$_2$/Si films. On the other hand, NiSi$_2$/Si films are much more disordered. Their spin-orbit scattering time is measured and compared with that in CoSi$_2$/Si films. 

The deposition and thermal annealing processes for the growth of TiSi$_2$/Si films and NiSi$_2$/Si films were similar to those for the fabrication of CoSi$_2$/Si films. The as-deposited Ti (Ni) films were annealed at 800$^\circ$C for 1 h in a high vacuum ($\approx$\,3\,$\times$\,$10^{-6}$ Torr) to form the C54 phase TiSi$_2$ (NiSi$_2$) on silicon. The lattice structure of NiSi$_2$ is identical to that of CoSi$_2$, namely, a centrosymmetric face-centered cubic (fcc) fluorite structure with $-$0.3\% lattice mismatch to Si \cite{Lambrecht1987}. Thus, NiSi$_2$ can easily form epitaxy on Si. The C54 phase TiSi$_2$ has a face-centered orthorhombic structure which consists of recursion of four-layer stacking \cite{Jeon1992}. The TiSi$_2$/Si films are polycrystalline, with an average grain size much smaller than that in CoSi$_2$/Si films. [We note that in addition to the low-resistivity C54 phase TiSi$_2$, there exists a C49 phase TiSi$_2$, which has a smaller grain size than that in the C54 phase. The resistivity of the C49 phase is much higher, with the residual resistivity $\rho_0$(C49) $\sim$\,100 $\mu\Omega$ cm $\gg \rho_0$(C54) \cite{Chiu2023,Mammoliti2002}. In this work, we shall focus on the C54 phase.]

\subsection{Electrical-transport measurements}

In all cases, the electrical transport measurements were carried out employing the standard four-probe technique. Linear Research Model LR400 and LR-700 ac resistance bridges were used. They supplied a low-level signal operating at 16 Hz for measuring sample resistance while avoiding electron overheating at low temperatures ($T$). An Oxford Heliox $^3$He cryostat equipped with a 2-T NbTi superconducting magnet or a standard $^4$He cryostat provided the low-$T$ environment. Several TiSi$_2$/Si films were measured in a BlueFors Model LD400 dilution refrigerator down to 50 mK.

\section{Normal-state electronic conduction properties}

In this section, we discuss the electrical-transport properties from resistivity $\rho (T)$ and the Hall effect measurements, relevant electronic parameters calculated from the measured residual resistivity ($\rho_0$) and carrier concentration ($n$), the spin-orbit scattering time ($\tau_{so}$) and electron dephasing length ($L_\varphi$) extracted from the weak-antilocalization (WAL) measurements, and low-frequency 1/$f$ noise in CoSi$_2$/Si films are discussed. The conduction properties of TiSi$_2$/Si films and NiSi$_2$/Si films are also measured and discussed for comparison where appropriate.   

\subsection{Electronic conduction properties}

\textbf{Temperature dependence of resistivity.} The electrical-transport properties of CoSi$_2$/Si(100) films and CoSi$_2$/Si(111) films reveal typical Boltzmann transport behavior. As $T$ decreases from room temperature, $\rho$ decreases with decreasing $T$, reaching a residual resistivity, $\rho_0 \equiv \rho$(4\,K), before the film undergoes a sharp superconducting transition. Figure \ref{fig3}(a) shows the $\rho (T)$ curves for three CoSi$_2$/Si(111) films, with $t$ = 10.5, 35 and 105 nm, as indicated. Figure \ref{fig3}(b) shows the $\rho (T)$ curves below 1.8 K for five CoSi$_2$/Si(100) films. Each film becomes superconducting at a temperature between $\approx$ 1.25 and 1.52 K, depending on $t$. The $\rho_0$ values of our thickest films are compatible with those of CoSi$_2$ single crystals \cite{Radermacher1993} and arc-melted bulk CoSi$_2$ (see below). For our highest quality films, the superconducting transition width, $\triangle T_c$, is as small as $\approx$\,5 mK. In most cases, $\triangle T_c \lesssim$ 15 mK can be achieved if the thermal annealing conditions are optimized. Here $\triangle T_c$ is the temperature difference between the temperature where $\rho$ drops to 0.9$\rho_0$ and the temperature where $\rho$ drops to 0.1$\rho_0$.

\begin{widetext}
\begin{center}
\begin{figure}[t!]
\includegraphics[width=1.0\textwidth]{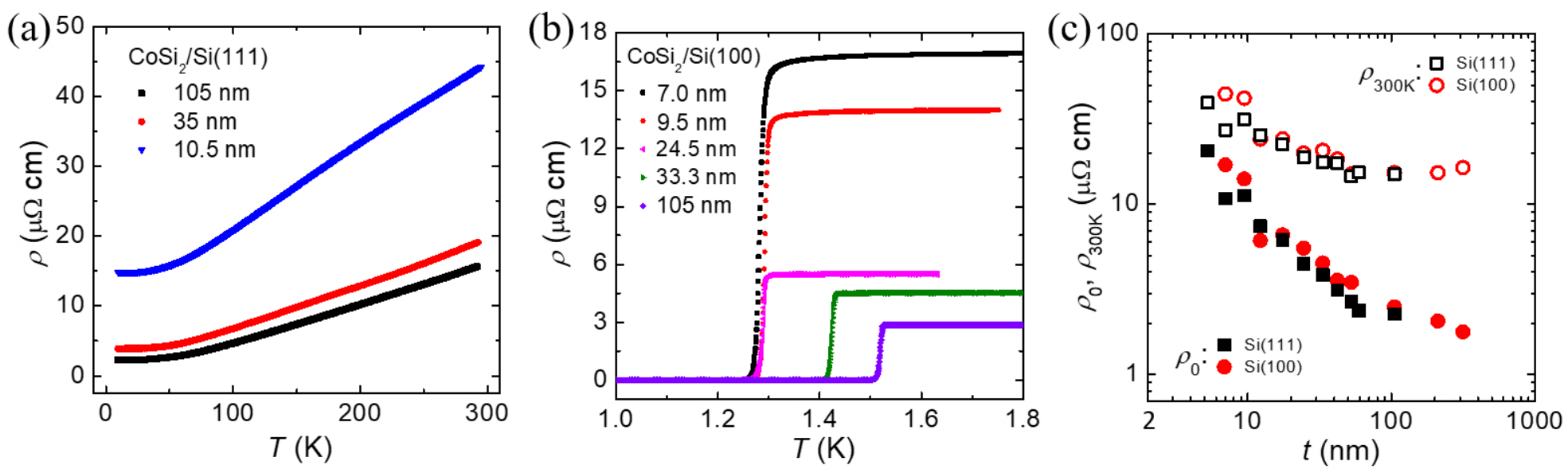} 
\caption{Variation of resistivity with temperature for CoSi$_2$/Si films.
(a) $\rho$ versus $T$ for three CoSi$_2$/Si(111) films with different $t$ values. (b) $\rho$ versus $T$ for five CoSi$_2$/Si(100) films below 1.8 K. (c) Room temperature resistivity $\rho_{\rm 300K}$ and residual resistivity $\rho_0$ versus log($t$) for a series of CoSi$_2$/Si(100) films and a series of CoSi$_2$/Si(111) films, as indicated.
}
\label{fig3}
\end{figure}  
\end{center}
\end{widetext}

Our results of metallic $\rho(T)$ characteristics align with theoretical studies of the electronic conduction properties of CoSi$_2$. Band structure calculations indicate that the CoSi$_2$ Fermi surface is relatively simple, consisting of three nested hole sheets centered at the Brillouin-zone origin. The broad low-lying Si 3$s$-3$p$ bands merge with the narrow Co 3$d$ bands to form the CoSi$_2$ Fermi surface \cite{Mattheiss1988}. Theoretical analysis also demonstrates that the measured $\rho(T)$ curve agrees reasonably well with a Bloch-Gr\"uneisen fit \cite{Allen1993}. 

\textbf{Residual resistivity versus film thickness $t$.}
The residual resistivity $\rho_0$ of CoSi$_2$/Si(100) films and a series of CoSi$_2$/Si(111) films have been measured and compared. Figure \ref{fig3}(c) shows the variation of the room temperature resistivity ($\rho_{\rm 300K}$) and  $\rho_0$ with log($t$). For the thickest film ($t$ = 315 nm) fabricated, a very low value of $\rho_0$ = 1.77 $\mu\Omega$ cm was obtained. This resistivity value is even lower than those of many elementary metals and metal alloys \cite{Kittel2005}, indicating one of the beneficial conduction attributes of CoSi$_2$/Si films. For example, these films can be used as interconnects in ICs \cite{Chen2004}. As $t$ decreases, $\rho_0$ progressively increases, as expected, due to increasing (partial specular) surface/interface scattering. For the films with $t \approx$ 7 nm, $\rho_0 \simeq 17$ and 11 $\mu\Omega$ cm for CoSi$_2$/Si(100) and CoSi$_2$/Si(111) films, respectively. For $t \gtrsim 10$ nm, the $\rho_0 (t)$ values are approximately the same for both series of films. Close inspection indicates that systematically, CoSi$_2$/Si(111) films have slightly lower $\rho_0 (t)$ values than CoSi$_2$/Si(100) films. Such information about the $\rho_0 (t)$ variation will be useful for the design of, \textit{e.g.}, CoSi$_2$/Si superconducting microwave resonators with desirable geometric inductance and kinetic inductance \cite{Zmuidzinas2012}. 

\textbf{Hall effect and classical magnetoresistance.}
The responsible carrier type and carrier concentration $n$ have been determined through the Hall effect measurements at liquid-helium temperatures and above $T_c$. The charge carriers in CoSi$_2$/Si films are found to be holes rather than electrons, in consistency with the band structure calculations \cite{Mattheiss1988,Newcombe1988}. Based on the free-electron gas model and assuming a single band, we obtain an average value $n \simeq$ (2.1$\pm$0.3)\,$\times$\,$10^{28}$ m$^{-3}$ for all films, independent of $t$ and the substrate orientation Si(100) or Si(111). With the values of $\rho_0$ and $n$ obtained, the elastic carrier/electron mean free time, $\tau_e = m^\ast/(ne^2\rho_0)$, for a given film can be calculated, by taking the effective carrier mass ($m^\ast$) to be the free electron mass ($m_e$) \cite{Newcombe1988}. From the $n$ value, the Fermi wavenumber $k_F = (3\pi^2 n)^{1/3} \approx$ 9\,$\times$\,$10^9$ m$^{-1}$ and the Fermi velocity $v_F = \hbar k_F/m^\ast \approx$ 1\,$\times$\,$10^6$ m/s can be calculated, where $\hbar$ is the reduced Planck constant. In practice, we obtain $\rho_0 l_e = 3\pi^2 \hbar/k_F^2e^2 \approx$ 1.6\,$\times$\,$10^{-15}$ $\Omega$ m$^2$ for CoSi$_2$/Si films, where $l_e = v_F \tau_e$ is the elastic carrier/electron mean free path. For example, for a $\rho_0$ = 5 $\mu\Omega$ cm CoSi$_2$/Si film, $\tau_e \simeq$ 3.4\,$\times$\,$10^{-14}$ s, $l_e \simeq$ 34 nm, the product $k_F \l_e \simeq$ 300, and the electron diffusion constant $D = v_F^2 \tau_e/3 \simeq$ 110 cm$^2$/s. Thus, these films fall in the weakly disordered regime where the quantum-interference phenomena, such as the weak-(anti)localization effect and universal conductance fluctuations \cite{Radermacher1993,JJLin2002}, should occur in the normal state at low $T$. 

\begin{figure}[t!]
\begin{center}
\includegraphics[width=0.48\textwidth]{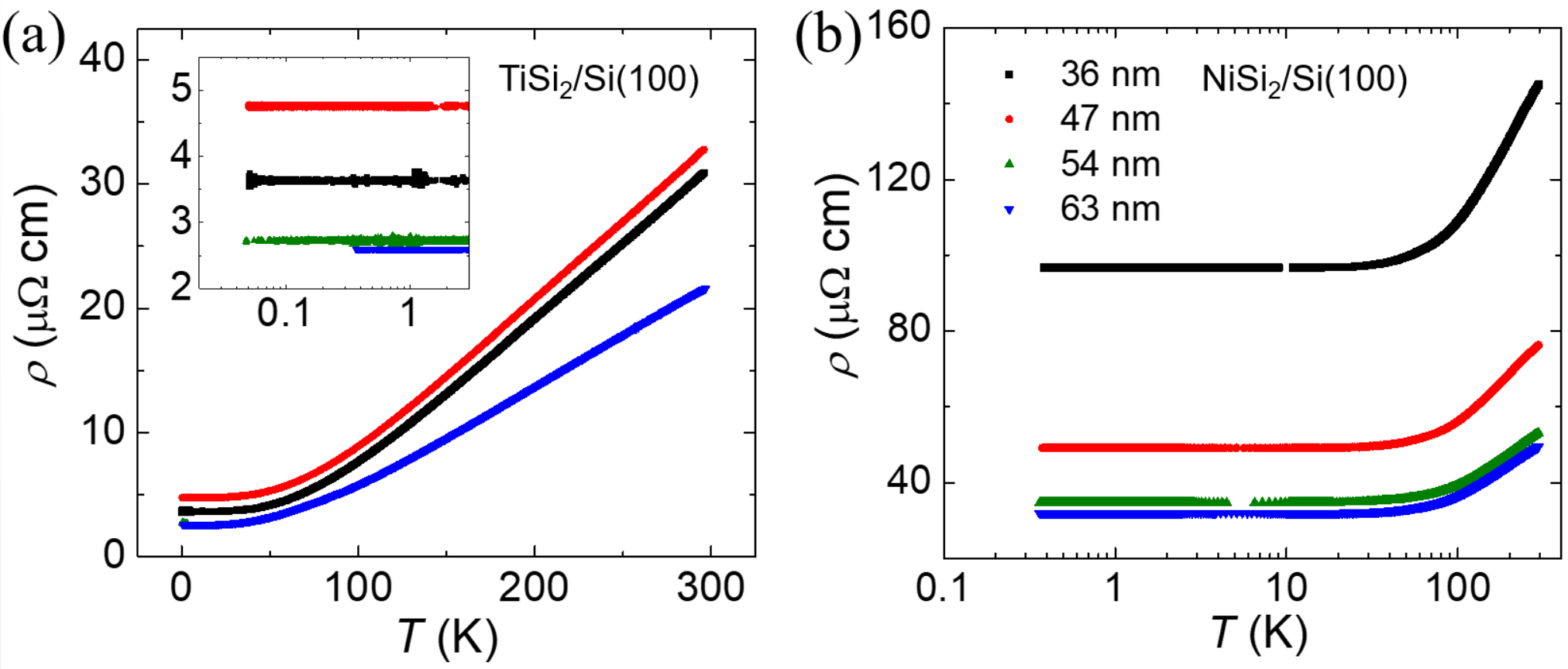} 
\caption{Variation of resistivity with temperature for TiSi$_2$/Si and NiSi$_2$/Si films.
(a) $\rho$ versus $T$ for four 125-nm thick TiSi$_2$/Si(100) films underwent different thermal annealing conditions and with different electron-beam lithographic patterned width ($W$): 800$^\circ$C, 1 h, $W$ = 250 $\mu$m (blue), 800$^\circ$C, 1 h, $W$ = 4.0 $\mu$m (green), 800$^\circ$C, 1 h, $W$ = 0.4 $\mu$m (black), and 750$^\circ$C, 1.5 h, $W$ = 0.4 $\mu$m (red). Inset: $\rho (T)$ below 3 K. (b) $\rho$ versus log($T$) for four NiSi$_2$/Si(100) films with differing thickness. Films were annealed at 800$^\circ$C for 1 h.    
}
\label{fig4}
\end{center}
\end{figure}

The classical MR due to the Lorentz force for several CoSi$_2$/Si films has been measured in perpendicular magnetic fields $|B| \leq$ 2 T and at low temperatures. The MR shows a parabolic dependence on $B$ (not shown), typical of a diffusive normal metal. Through the expression $\triangle R(B)/R(0) = [R(B)-R(0)]/R(0) = (e\tau_e B/m^\ast)^2$, one can also calculate the $\tau_e$ value. The inferred $\tau_e$ values are about a factor of $\sim$\,1.7 longer than those calculated from the measured $\rho_0$ and $n$ values mentioned above. This can be (partly) ascribed to the fact that a perpendicular $B$ field mainly probes the motion of charge carriers in the film plane. The estimates extracted from the $\rho_0$ and $n$ (\textit{i.e.}, the Hall effect) method and the classical MR method are based on the free-electron gas single-band model, so such a discrepancy is not unexpected. In the present work, the relevant electronic parameters are based on $n$  inferred from the Hall effect \cite{Note1}. 

\textbf{Resistivity of TiSi$_2$/Si and NiSi$_2$/Si films.}
TiSi$_2$ and NiSi$_2$ are nonmagnetic metals \cite{Jeon1992,Chen2004}. Figure \ref{fig4}(a) and \ref{fig4}(b) show the resistivity as a function of temperature for four TiSi$_2$/Si films and four NiSi$_2$/Si films, respectively. The two figures indicate that both disilicides have $\rho(T)$ monotonically decreasing with decreasing $T$, reaching a residual resistivity at low temperatures. The inset of Fig. \ref{fig4}(a) clearly shows that the TiSi$_2$/Si films remain non-superconducting down to at least 50 mK. Thus, this film material can serve as an excellent counterpart for fabricating transparent CoSi$_2$/TiSi$_2$ S/N heterojunctions for the phase-sensitive conductance spectroscopy studies to probe the unconventional pairing symmetry, as emphasized above. Figure \ref{fig4}(b) shows $\rho$ versus log($T$) for four NiSi$_2$/Si films with differing thickness. We notice that these films have $\rho_0$ values about one order of magnitude larger than those in CoSi$_2$/Si and TiSi$_2$/Si films, indicating that they are relatively disordered. Thus, they are convenient for the measurements of spin-orbit scattering time ($\tau_{so}$) through the quantum-interference effects. The log($T$) scale of the abscissa in Fig. \ref{fig4}(b) makes it clear to see that NiSi$_2$/Si films do not undergo a superconducting transition at least down to 0.3 K.  

Our estimates of the relevant electronic parameters for the C54 phase TiSi$_2$/Si films are as follows: $n \approx$ 3\,$\times$\,$10^{29}$ m$^{-3}$ \cite{Mammoliti2002}, $m^\ast \approx m_e$, $k_F \approx$ 2\,$\times$\,$10^{10}$ m$^{-1}$, $v_F \approx$ 2\,$\times$\,$10^{6}$ m/s, and the product $\rho_0 l_e \approx$ 2.8\,$\times$\,$10^{-16}$ $\Omega$ m$^2$. Our corresponding estimates for NiSi$_2$/Si films are: $n \approx$ 1.7\,$\times$\,$10^{28}$ m$^{-3}$ (from the Hall effect measurements), $m^\ast \approx m_e$, $k_F \approx$ 8\,$\times$\,$10^{9}$ m$^{-1}$, $v_F \approx$ 9\,$\times$\,$10^{5}$ m/s, and the product $\rho_0 l_e \approx$ 2\,$\times$\,$10^{-15}$ $\Omega$ m$^2$.

\subsection{Spin-orbit scattering time}

The spin-orbit coupling  has become of fundamental importance in condensed matter and quantum materials science. It plays a vital role in low-dimensional systems and heterostructures and can be relevant for producing unconventional superconducting states \cite{Sau2010,Tokatly2014,Stewart2017}. The ability to manipulate quantum states through SOC engineering may have future applications in spintronics and quantum-information technology \cite{Manchon2015}. 
In diffusive metals, the strength of SOC increases with increasing atomic number ($Z$) and growing degree of disorder, i.e., the spin-orbit scattering rate is predicted to obey the relation $\tau_{so}^{-1} \simeq (\alpha Z)^4/\tau_e$ \cite{Abrikosov1962}, where $\alpha \simeq 1/137$ is the fine-structure constant. This conventional wisdom suggests that the SOC is strong in disordered metals containing heavy impurities such as Au, Pt and Bi. 
In CoSi$_2$/Si films, as will be discussed below, the SOC is large but the films are comprised of comparatively light elements. Thus, an extrinsic mechanism is required to account for the large observed SOC. From the low-velocity limit of the Dirac equations, one has for  the SOC coupling term 
$H_{\mbox{\tiny SOC}}=\frac{[\nabla V(\vec{r})\times \vec{k}]\cdot \vec{S}}{2m_e c^2}$, where $V$ is the electron potential, $\vec{k}$ its momentum, and $\vec{S}$ describes the spin degree of freedom. Thus, if the potential gradient at the interface between CoSi$_2$ and Si in CoSi$_2$/Si films is large, a sizeable SOC interaction is generated. The large SOC in CoSi$_2$/Si films has thus been attributed to the Rashba SOC \cite{Mishra.21}.

\begin{widetext}
\begin{center}
\begin{figure}[t!]
\centering
\includegraphics[width=0.85\textwidth]{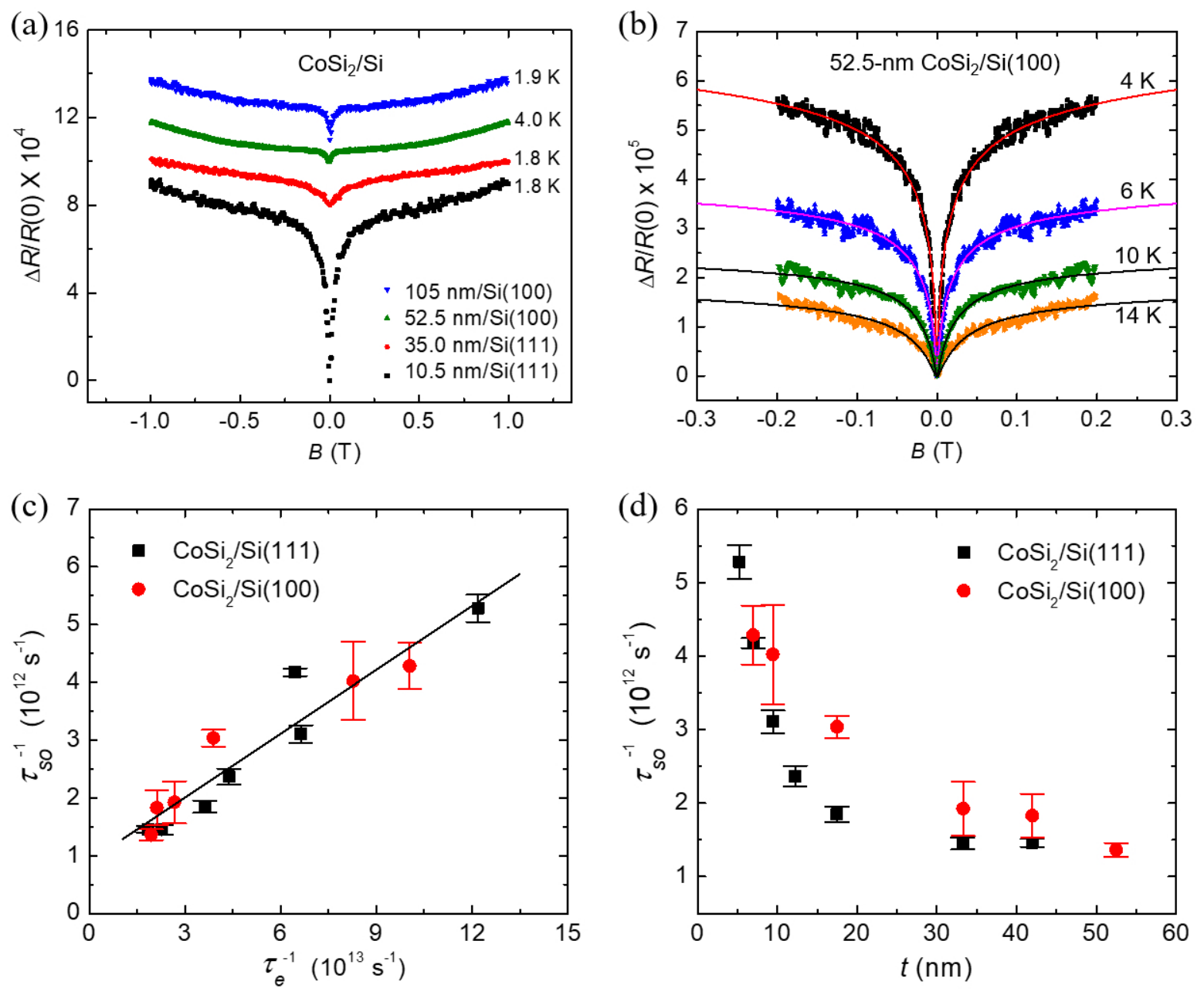} 
\caption{Weak-antilocalization MR and spin-orbit scattering rate of CoSi$_2$/Si films.
(a) Normalized MR, $\triangle R(B)/R(0)$, for four CoSi$_2$/Si films in perpendicular magnetic fields. (b) $\triangle R(B)/R(0)$ in low $B$ fields for a 52.5-nm thick CoSi$_2$/Si(100) film at four $T$ values. The solid curves are least-squares-fit to the WAL theory predictions. (c) $\tau_{so}^{-1}$ versus $\tau_e^{-1}$ for a series of CoSi$_2$/Si(100) films and a series of CoSi$_2$/Si(111) films. The straight solid line is a linear fit. (d) $\tau_{so}^{-1}$ as a functions of $t$.
}
\label{fig5}
\end{figure}
\end{center}
\end{widetext}

The $\tau_{so}^{-1}$ values of a series of CoSi$_2$/Si(100) films and a series of CoSi$_2$/Si(111) films have been quantitatively extracted from the quantum-interference weak-antilocalization (WAL) studies. It is well established that in the strong SOC limit $\tau_{so}^{-1} > \tau_\varphi^{-1}$, where $\tau_\varphi$ is the electron dephasing time, the WAL effect results in a positive MR in low $B$ fields in a weakly disordered metal. On the other hand, in the opposite limit of weak SOC ($\tau_{so}^{-1} < \tau_\varphi^{-1}$), a negative MR in low $B$ fields is expected from the weak-localization effect \cite{JJLin2002}.  

Figure \ref{fig5}(a) shows the normalized MR, $\triangle R(B)/R(0) = [R(B)-R(0)]/R(0)$, for four CoSi$_2$/Si films measured in the normal state and in perpendicular $B$ fields, \textit{i.e.}, $B$ is applied perpendicular to the film plane. In the relatively large $B$ field regime ($B \gg B_\varphi = \hbar/4eD\tau_\varphi^2$), a parabolic MR background resulting from the classical Lorentz force is seen. What is more important is the notable positive MR in the low $B$ field regime ($B \lesssim B_\varphi$), which manifests the WAL effect and provides direct evidence for strong SOC in CoSi$_2$/Si films. Figure \ref{fig5}(b) shows the low-$B$ field MR for a 52.5-nm thick CoSi$_2$/Si(100) film measured at four $T$ values. The solid curves are least-squares fits to the theoretical predictions of Hikami \textit{et al.} \cite{Hikami1980}, taking the superconducting fluctuation effect into account \cite{Wu1994}. Good agreement between theory and experiment is obtained. Thus, the temperature-dependent $\tau_\varphi (T)$ and the temperature insensitive $\tau_{so}$ can be reliably extracted for each film.

\begin{center}
\begin{figure}[t!]
\centering
\includegraphics[width=0.48\textwidth]{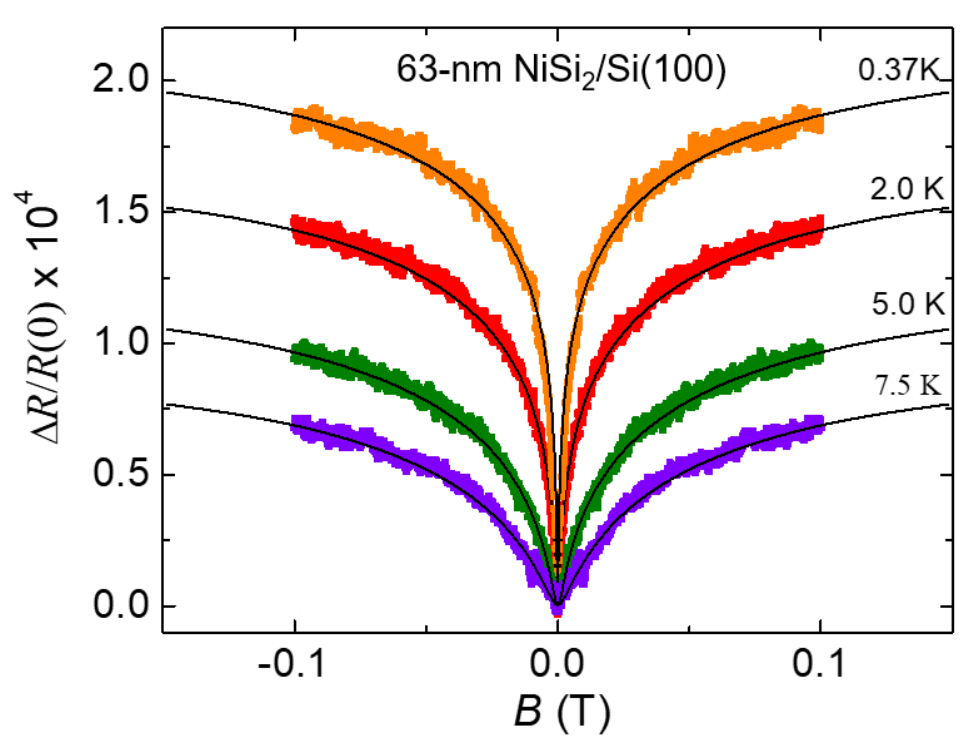} 
\caption{Normalized MR, $\triangle R(B)/R(0)$, for a 63-nm thick NiSi$_2$/Si(100) film in perpendicular $B$ fields and at four $T$ values. The solid curves are the WAL theory predictions. 
}
\label{fig6}
\end{figure}
\end{center}

Figure \ref{fig5}(c) shows the variation of the extracted $\tau_{so}^{-1}$ with elastic electron scattering rate $\tau_e^{-1}$ for a series of CoSi$_2$/Si(100) films and a series of CoSi$_2$/Si(111) films. This figure indicates that $\tau_{so}$ is extremely short, falling between 0.2 and 0.8 ps, depending on the $\tau_e^{-1}$ value. Inspection of Fig. \ref{fig5}(c) reveals that, for a given $\tau_e^{-1}$ value, the $\tau_{so}^{-1}$ value is essentially the same for films grown on Si(100) and Si(111) substrates. Moreover, the figure reveals an approximately linear variation $\tau_{so}^{-1} \propto \tau_e^{-1}$, suggesting that the SOC is leading to spin relaxation through the Elliott-Yafet process \cite{Fabian1999}. The underlying origin for the strong spin-orbit interaction is most likely the Rashba SOC mechanism induced by the broken inversion symmetry at the sharp and strain-free CoSi$_2$/Si interface as suggested in Ref.~\cite{Mishra.21}. In fact, in our CoSi$_2$/Si films, the ratio $\l_e/t$ gradually increases from $\approx$\,1 to $\approx$\,2.2 as $t$ decreases from 105 nm to 5 nm (not shown), suggesting that surface and interface scattering dominates over the bulk defect scattering. This is illustrated  in Figure \ref{fig5}(d) which shows a plot of $\tau_{so}^{-1}$ versus $t$. An increase of $\tau_{so}^{-1}$ with decreasing $t$, \textit{i.e.}, increasing CoSi$_2$/Si interface scattering, especially at $t \lesssim$ 20 nm is evident. This finding further supports the interface-induced Rashba SOC scenario \cite{Mishra.21,Chiu2023}. Further experimental and theoretical studies to address this issue are desirable.

\begin{table}
\caption{\label{table1}
Relevant parameters for spin-orbit scattering time $\tau_{so}$ in CoSi$_2$/Si and NiSi$_2$/Si films with $t \geq$ 10 nm. Values for representative Al films (taken from \cite{Santhanam1987}), Cu and Au films (taken from \cite{Pierre2003}), and Ti$_{73-x}$Al$_{27}$Sn$_x$ (0\,$\leq$\,$x$\,$\leq$\,5) alloys (taken from \cite{Hsu1999}) are listed for comparison. $Z$ is the atomic number, $\rho_0$ the residual resistivity, and $\tau_e$ the elastic mean free time. For the disilicides, the listed $Z$ value is the average value $[Z($Cu,Ni)+$Z({\rm Si})]/3$. For the Ti-Al-Sn alloy, the listed $Z$ value is an average value over the composition.
}
\begin{ruledtabular}
\begin{tabular}{lrlll}
Films & $Z$ & $\rho_0$\,($\mu\Omega$\,cm) & $\tau_e$\,(fs) & $\tau_{so}$\,(ps) \\ 
\hline 
CoSi$_2$/Si & $\approx$\ 18 & $\approx$ 2--10 & $\approx$ 8--55 & $\approx$ 0.2--0.8 \\
NiSi$_2$/Si & $\approx$ 19 & $\approx$ 32--64 & $\approx$ 3.5--6.4 & $\approx$ 10--30 \\
Al & 13 & $\approx$ 1.3--5.9 & $\approx$ 5--20 & $\approx$ 20--100 \\
Cu & 29 & $\approx$ 1.6--5.0 & $\approx$ 6--19 & $\approx$ 10--30 \\
Au & 79 & $\approx$ 2.5 & $\approx$ 20 & $\approx$ 0.5 \\
Ti$_{73-x}$Al$_{27}$Sn$_x$ & $\approx$ 20 & $\approx$ 225 & $\approx$ 0.1 & $\approx$ 3--20\\ 
\end{tabular}
\end{ruledtabular}
\end{table}

For comparison, the spin-orbit scattering time for four NiSi$_2$/Si(111) films with $t$ in the range 36 to 63 nm, corresponding to $\rho_0$ varying from 64 down to 32 $\mu\Omega$ cm, has also been measured through the WAL effect. Figure \ref{fig6} shows the results for a representative film. We obtain $\tau_{so} \approx$ (10--30) ps, being more than one order of magnitude longer than that in CoSi$_2$/Si films, even though these NiSi$_2$/Si films are more disordered than CoSi$_2$/Si films, see Table \ref{table1}. 

Table \ref{table1} also listed the $\tau_{so}$ values for several typical polycrystalline Al films \cite{Santhanam1987}, Cu and Au films \cite{Pierre2003} and Ti$_{73-x}$Al$_{27}$Sn$_x$ (0\,$\leq$\,$x$\,$\leq$\,5) alloys \cite{Hsu1999} taken from the literature. The $\rho_0$ and $\tau_e$ values of these elementary metal and disilicide (except NiSi$_2$/Si) films listed in Table \ref{table1} are on the same order of magnitude. However, the $\tau_{so}$ value of CoSi$_2$/Si films is much shorter than that in the other (except Au) films. That is, $\tau_{so}$(CoSi$_2$/Si) is about two orders of magnitude shorter than those in Al, Cu and NiSi$_2$/Si films. Surprisingly, even in the very high $\rho_0$ ($\simeq$ 225 $\mu\Omega$ cm) Ti$_{73-x}$Al$_{27}$Sn$_x$ alloys where $l_e$ approaches the interatomic spacing, its $\tau_{so}$ value is still more than one order of magnitude larger than that in CoSi$_2$/Si films. It is also intriguing that the SOC rate in CoSi$_2$/Si films is comparable with that in a heavy Au film, especially considering that $Z$(Au) is much larger than $Z$(CoSi$_2$/Si). The above results strongly point out that, in many cases, the simple relation $\tau_{so}^{-1} \propto Z^4/\tau_e$ does not apply across different metals and alloys containing (heavy) impurities \cite{Bergmann1992}.

While $\tau_{so}^{-1}$(CoSi$_2$/Si) $\gg$ $\tau_{so}^{-1}$(Al), the two materials have very similar $T_c$ and $\triangle_0$ values. In other words, the ratios of the SOC energy ($\hbar/\tau_{so}$) to $\triangle_0$ are significantly different in CoSi$_2$/Si films and Al films. In the former, $(\hbar/\tau_{so})/\triangle_0 \sim$ 15, and in the latter, $(\hbar/\tau_{so})/\triangle_0 \sim$ 0.1. For the (seven) Al films listed in Table 1, $T_c \approx$ 1.4 K. $\tau_{so} \approx$ 20-100 ps. The ratio $(\hbar/\tau_{so})/\triangle_0 \approx$ 0.03--0.2. For the CoSi$_2$/Si film with the shortest $\tau_{so} \approx$ 0.2 ps and the lowest $T_c \approx$ 1.15 K, $\hbar/\tau_{so} \approx$ 3.5 meV, and $(\hbar/\tau_{so})/\triangle_0 \approx$ 20. In Ref. \cite{Chiu.21} we used the classical MR to evaluate the electronic parameter. In this work, we have measured the Hall effect and started with the evaluated $n$ value (through a single-band free-electron-gas model) to calculate the electronic parameters, see subsection III.A. This leads to the ratio $(\hbar/\tau_{so})/\triangle_0$ a factor $\sim$\,1.5 smaller than that ($\approx$\,30) in Ref. \cite{Chiu.21}.

\textbf{Generation of two-component superconductivity.} The CoSi$_2$ band structure may be substantially modified in CoSi$_2$/Si films due to the confined geometry and changes in the chemical environment. On top of these changes, the interface generates a SOC term 
$H_{\mbox{\tiny SOC}}=\frac{[\nabla V(\vec{r})\times \vec{k}]\cdot \vec{S}}{2m_e c^2} \sim \vec{\mathcal{A}}(k)\cdot \vec{\sigma}$ with $|\vec{\mathcal{A}}(k)|=1$, as briefly discussed above. Here, $\vec{\mathcal{A}}(k)=(k_y,-k_x,0)$,
and $\vec{\sigma}$ is a vector of Pauli spin matrices ($\sigma_x$,$\sigma_y$,$\sigma_z$). This leads to splitting the otherwise spin-degenerate bands into two helical bands. The SOC energy is generally minimal compared to the Fermi energy but, as it turns out, is large in CoSi$_2$/Si films compared to the characteristic energy $\Delta_0$ of the superconducting state. Revisiting the BCS pairing problem in the presence of this SOC term, one obtains a gap structure $\hat{\Delta}=\big( \Delta_s+\Delta_t\, \vec{\mathcal{A}}\cdot \vec{\sigma} \big)i\sigma_y$ \cite{Mineev.94}. If one compares with the general gap structure $\hat{\Delta}=\Delta_s i\sigma_y+\Delta_t  \vec{d}\cdot  \vec{\sigma} i \sigma_y$, it becomes clear that the SOC interaction leads to an order parameter that possesses both spin-singlet ($\Delta_s$) and spin-triplet ($\Delta_t$) component. The SOC vector $\vec{\mathcal{A}}$ acts as the spin vector $\vec{d}$-vector \cite{Annunziata2012,Mishra.21}. The effect of a superconductor with strong SOC in place of the $s$-wave or $p$-wave superconductor of a T-shaped proximity structure was analyzed in Ref.~\cite{Mishra.21}. As it turns out, the resulting conductance spectrum can feature a sharp peak as in Fig.~\ref{fig1}(d) for a wide parameter range where the spin-triplet component dominates. The conductance spectrum will interpolate between the spin-singlet and spin-triplet cases illustrated in Fig.~\ref{fig1}(c) as a function of the relative weight of the two components \cite{Mishra.21}. 

\subsection{Electron dephasing length} 

\begin{center}
\begin{figure}[t!]
\centering
\includegraphics[width=0.48\textwidth]{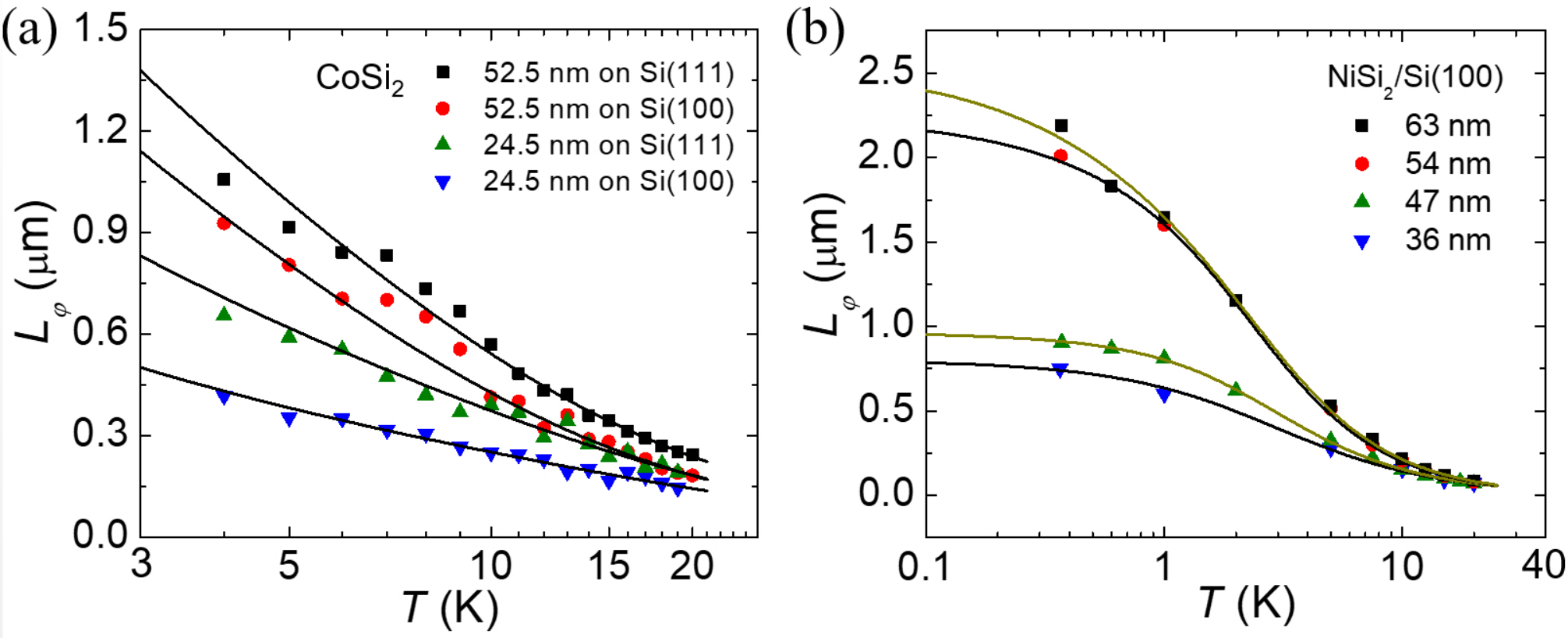} 
\caption{Electron dephasing length 
$L_\varphi = \sqrt{D \tau_\varphi}$ as a function of log($T$) for (a) four CoSi$_2$/Si films, and (b) four NiSi$_2$/Si films. The solid curves in (a) and (b) are least-squares  fits with the total electron dephasing rate $\tau_\varphi^{-1}$ as an adjusting parameter (see text). 
}
\label{fig7}
\end{figure}
\end{center}

In addition to $\tau_{so}$, the electron dephasing time $\tau_\varphi$ can also be extracted from the WAL measurements. Figure \ref{fig7}(a) shows the electron dephasing length $L_\varphi = \sqrt{D \tau_\varphi}$ as a function of $T$ for four CoSi$_2$/Si films. The WAL effect is measured only down to 4 K to minimize (but still not totally eliminate) the superconducting-fluctuation-effect induced positive MR \cite{Wu1994}. The solid curves are least-squares fits with the total dephasing rate expressed by $\tau_\varphi^{-1} = \tau_{ee}^{-1} + \tau_{ep}^{-1} = A_{ee}T + A_{ep}T^3$ \cite{JJLin2002}, where the first term on the right-hand side of the equation is the two-dimensional electron-electron scattering rate, and the second term is the electron-phonon scattering rate. For the two 52.5-nm thick films with $\rho_0 \approx$ 3.0 $\mu\Omega$ cm, we obtain approximate values of $A_{ee}\approx$ 3.6\,$\times$\,$10^9$ s$^{-1}$\,K$^{-1}$ and $A_{ep}\approx$ 4.6\,$\times$\,$10^7$ s$^{-1}$\,K$^{-3}$. For the two 24.5-nm thick films with $\rho_0 \approx$ 4.5 $\mu\Omega$ cm, we obtain approximate values of $A_{ee}\approx$ 1.0\,$\times$\,$10^{10}$ s$^{-1}$\,K$^{-1}$ and $A_{ep}\approx$ 3.3\,$\times$\,$10^7$ s$^{-1}$\,K$^{-3}$. Here the  $A_{ep}$ values are about a factor of $\sim$\,2.5 higher than that in those Al films studied in Ref. \cite{Santhanam1987}. The information about $\tau_{ep}^{-1}$ will be useful for understanding the quasiparticle dissipation problem in the superconducting devices made of CoSi$_2$/Si films. 

We note that Fig. \ref{fig7}(a) reveals a long $L_\varphi$ exceeding 1 $\mu$m already at 4 K for films with $t >$ 50 nm. We also would like to point out that, taken together, (1) the long $L_\varphi$, (2) a high $T_c$ accompanied by a sharp superconducting transition [Fig. \ref{fig2}(b)], and (3) the SQUID magnetization measurement results \cite{Chiu.21}, complementarily suggest that our CoSi$_2$/Si films are nonmagnetic. For comparison, we note that our measured $L_\varphi$ values are much longer than those previously reported by DiTusa \textit{et al.} in their CoSi$_2$ epitaxial films, where the authors ascribed their short $L_\varphi$ to the contamination of magnetic impurities and possibly small departure from stoichiometry at the film/substrate interface \cite{DiTusa1990}.

Figure \ref{fig7}(b) shows $L_\varphi$ as a function of $T$ for four NiSi$_2$/Si(100) films. Because these films do not become superconducting, we can measure the WAL MR and extract $L_\varphi$ down to sub-kelvin temperatures. The solid curves are least-squares fits with the total dephasing rate $\tau_\varphi^{-1} = A_0 + A_{ee}T + A_{ep}T^p$ \cite{JJLin2002}, where the additional term on the right-hand side of the equation is a constant, called the ``saturated dephasing rate" and given by $A_0 = \tau_\varphi^{-1}(T$\,$\rightarrow$\,0), and $p$ is a temperature exponent for the electron-phonon scattering. For example, for the 63-nm thick film with $\rho_0 \approx$ 32 $\mu\Omega$ cm, we obtain $A_0 \approx$ 2.9\,$\times$\,$10^8$ s$^{-1}$, $A_{ee}\approx$ 3.3\,$\times$\,$10^8$ s$^{-1}$\,K$^{-1}$ and $A_{ep}\approx$ 6.0\,$\times$\,$10^8$ s$^{-1}$\,K$^{-p}$, with $p \approx$ 2.8. Figure~\ref{fig7}(b) reveals a $L_\varphi$ reaching 2 $\mu$m at low $T$, suggesting that this disilicide can be an appealing material candidate for making quantum-interference devices. Our extracted $L_\varphi$ values are in good accord with those previously reported by Matsui \textit{et al.} in single-crystal NiSi$_2$ films \cite{Matsui1990}.

\subsection{Low frequency 1/$f$ noise}

The low-frequency (flicker) 1/$f$ noise in a conductor is empirically described by the Hooge relation $S_V = \gamma V_s^2/N_c f^\beta + S_V^0$,  where $S_V$ is the measured voltage noise power spectrum density (PSD), $\gamma$ is a dimensionless parameter characterizing the magnitude of the noise, $V_s$ is the bias voltage drop on the sample, $N_c$ is the total number of charge carriers in the conductor, $f$ is the frequency, and the exponent $\beta \approx 1$ for a wide variety of conductors \cite{Dutta1981}. $S_V^0$ is the background noise of the measurement circuit,  which is limited by the Johnson-Nyquist noise and the input noise $\approx 1.7 \times 10^{-17}$ V$^2$/Hz of our preamplifier (Stanford Research Systems model SR560). The underlying origin of the 1/$f$ noise is usually modeled by the two-level systems which are taken to be dynamical structural defects, such as a small group of moving atoms, oxygen vacancies, dangling bonds, nanometer-sized grains, etc. \cite{Yeh2017,Yeh2018} For any practical nanoelectronic and superconducting devices to achieve the ultimate performance, it is highly desired that the 1/$f$ noise level of the device be as low as possible.

CoSi$_2$/Si films have the advantage that the calcium fluoride (CaF$_2$) crystal structure of CoSi$_2$ has strong covalent bonding. Moreover, as mentioned, the lattice mismatch between CoSi$_2$ and silicon is minimal. This small lattice mismatch makes favorably the epitaxial growth of CoSi$_2$ on both Si(100) and Si(111) substrates, resulting in very few dangling bonds at the interface. Thus, CoSi$_2$/Si films are fairly stable in the ambient and expected to have an ultralow level of 1/$f$ noise \cite{Chiu.17}.

\begin{center}
\begin{figure}[t!]
\centering
\includegraphics[width=0.4\textwidth]{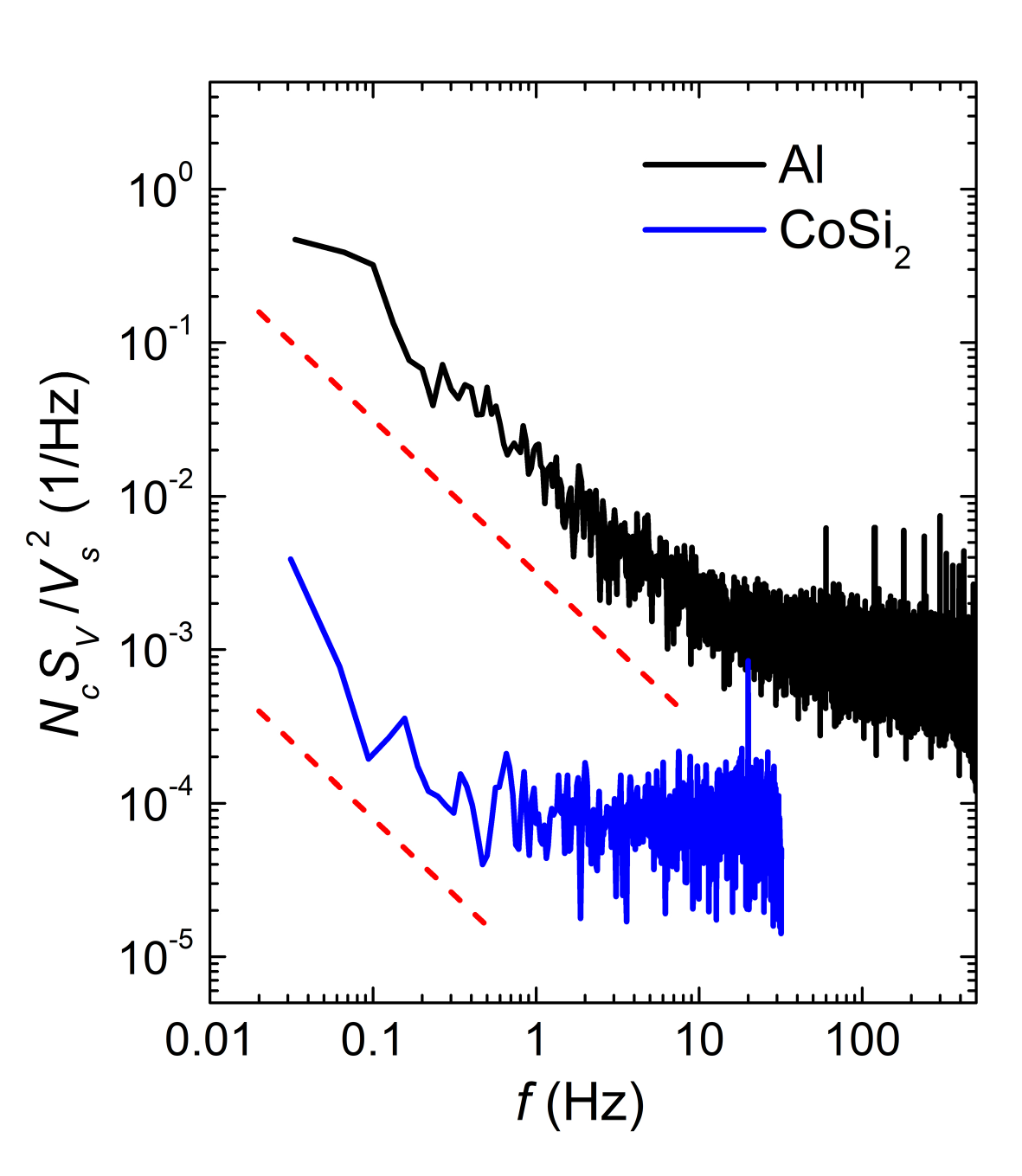} 
\caption{Normalized low-frequency noise $N_cS_V/V_s^2$ versus $f$ of a 105-nm thick CoSi$_2$/Si(111) film with $\rho$(300\,K) = 17 $\mu\Omega$ cm and a 30-nm thick Al film with $\rho$(300\,K) = 5.5 $\mu\Omega$ cm at $T$\,=\,300 K. The bias voltage was $\approx$\,146 mV ($\approx$\,131 mV) for the CoSi$_2$/Si (Al) film. The straight dashed lines are drawn proportional to $f^{-1}$ and is guide to the eye. 
}
\label{fig8}
\end{figure}
\end{center}

Figure \ref{fig8} shows the variation of the normalized low-frequency noise $N_c S_V/V_s^2$ with $f$ for a 105-nm thick CoSi$_2$/Si(111) film and a 30-nm thick polycrystalline Al film at $T$\,=\,300 K. The Al film was deposited via electron-gun evaporation. A 1/$f^\beta$ dependence with $\beta \simeq$ 1 (as indicated by the straight dashed lines) is seen already at $f \lesssim$ 30 Hz in the Al film, while it only appears at very low frequencies $f \lesssim$ 0.3 Hz in the CoSi$_2$/Si(111) film. This is a direct manifestation of the fact that the excess 1/$f$ noise is minimal and buried in the background noise $S_V^0$ until at very low frequencies. We have measured the $S_V \propto V_s^2$ dependence in both Al and CoSi$_2$/Si(111) films and determined the $\gamma$ values from the slope. The extracted $\gamma$(300\,K) $\approx$ (5$\pm$3)\,$\times$\,$10^{-6}$ for the CoSi$_2$/Si(111) film is very small and in line with our previous result of $\gamma$(150\,K) $\approx$ 3\,$\times$\,$10^{-6}$ in another 105-nm thick CoSi$_2$/Si(100) film \cite{Chiu.17}. This value is three orders of magnitude lower than the extracted $\gamma$(300\,K) $\approx$ (5$\pm$2)\,$\times$\,$10^{-3}$ in the polycrystalline Al film (Fig. \ref{fig8}) and two orders of magnitude lower than that in single-crystalline Al films grown on sapphire substrate \cite{Van1998,Scofield1985}. We note that, for typical metals and alloys, $\gamma\approx$ $10^{-4}$--$10^{-2}$ \cite{Yeh2017,Yeh2018}. Our observation of ultralow 1/$f$ noise suggests that CoSi$_2$/Si films have a high potential for use as building blocks for superconducting circuits and quantum devices.

\section{Superconducting properties}

In this section, we discuss the variation of superconducting properties with the thickness of CoSi$_2$/Si films. The superconducting transition temperature $T_c$ as well as the upper critical fields $B_{c2,\perp}$ and $B_{c2,\parallel}$ measured in magnetic fields applied perpendicular and parallel to the film plane, respectively, will be discussed. The Ginzburg-Landau coherence length $\xi_{\rm GL}(T)$ in the film plane can be inferred from the measured $B_{c2,\perp}(T)$ value. The penetration depth $\lambda (T$\,$\rightarrow$\,0) in the BCS theory is calculated. We will remark on the potential applications of CoSi$_2$/Si films as superconducting microwave resonators. The critical current density of films as well as the superconducting properties of polycrystalline CoSi$_2$ bulks will be briefly discussed.

\begin{figure}[t!]
\centering
\includegraphics[width= 0.45\textwidth]{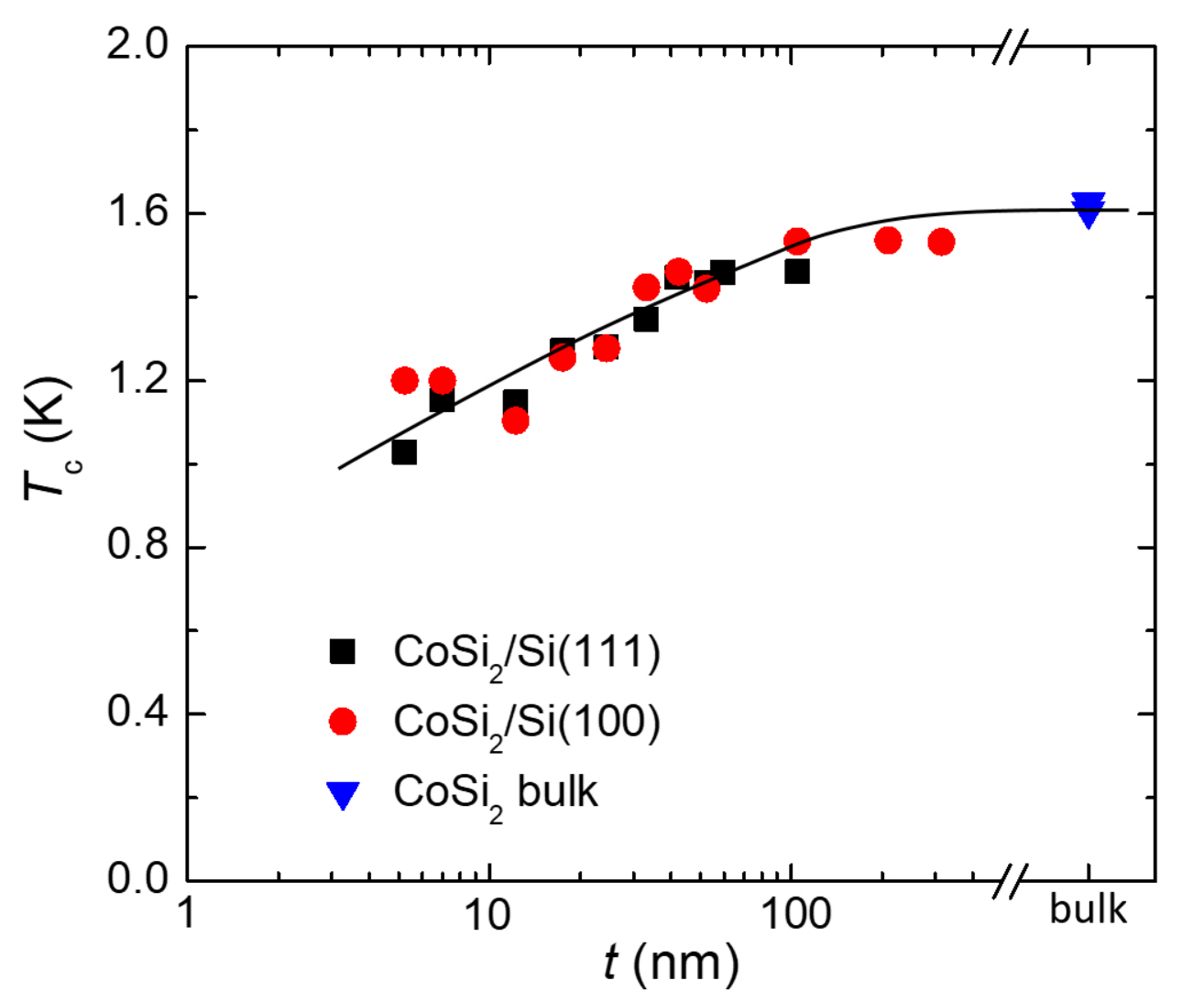} 
\caption{Superconducting transition temperature $T_c$ as a function of $t$ for a series of CoSi$_2$/Si(100) films and a series of CoSi$_2$/Si(111) films. The solid curve is a guide to the eye. The two blue down triangles represent the $T_c$ values for arc-melted polycrystalline CoSi$_2$ bulks (cf. Fig. \ref{fig12}).
}
\label{fig9}
\end{figure}

\subsection{Variation of superconducting transition temperature with film thickness}

The superconducting transition temperature $T_c$ of a series of CoSi$_2$/Si films grown on Si(100) substrates and a series of CoSi$_2$/Si films grown on Si(111) substrates have been measured. The variation of $T_c$ with CoSi$_2$/Si film thickness $t$ is shown in Fig. \ref{fig9}. This figure indicates that the $T_c$ value for a given $t$ is essentially the same for both series of films, independent of the Si(100) and Si(111) substrate direction. Moreover, the $T_c$ value ($\simeq$ 1.45$\pm$0.1 K) remains essentially unchanged for films with $t \gtrsim 35$ nm. As $t$ decreases from 35 nm down to 7 nm, the $T_c$ value gradually decreases from $\approx$\,1.4 K to $\approx$\,1.2 K. The thinnest films which still become superconducting have a thickness $t \simeq$ 5 nm and a transition temperature $T_c \simeq$ (1.0--1.2) K. We have also fabricated a 3.5-nm thick CoSi$_2$/Si(111) film, which has a fairly high residual resistivity $\rho_0 = 56.3$ $\mu \Omega$ cm and a superconducting onset temperature $\approx$\,0.6 K. A zero-resistivity state was not reached down to 0.36 K. For comparison, we note that in the early work by DiTusa \textit{et al.} \cite{DiTusa1990}, the $T_c$ value of their 20-nm (12-nm) thick CoSi$_2$ epitaxial films prepared using ultrahigh vacuum techniques already dropped to 1.06 K ($<$ 0.57 K). Our higher $T_c$ values compared with theirs (of similar $t$) suggest 
the reliability of our fabrication method and the high quality of our as-grown films.   

With the residual resistivity $\rho_0$ (equivalently, the sheet resistance $R_\square = \rho_0/t$) and the $T_c$ value available as a function of $t$ [Fig. \ref{fig3}(c) and Fig. \ref{fig9}], one may design and grow patterned CoSi$_2$/Si films by modifying the thickness and geometry to adjust, \textit{e.g.}, the kinetic inductance ($L_K$) of a microwave resonator for a specific superconducting circuit. Approximately, $L_K \propto (L/W)(R_\Box /\triangle_0)$ as $T \rightarrow 0$ \cite{Zmuidzinas2012, Mukhanova2023}, where $L$ and $W$ are the length and width of the resonator, respectively.

\subsection{Upper critical fields}

\begin{figure}[t!]
\centering
\includegraphics[width= 0.48\textwidth]{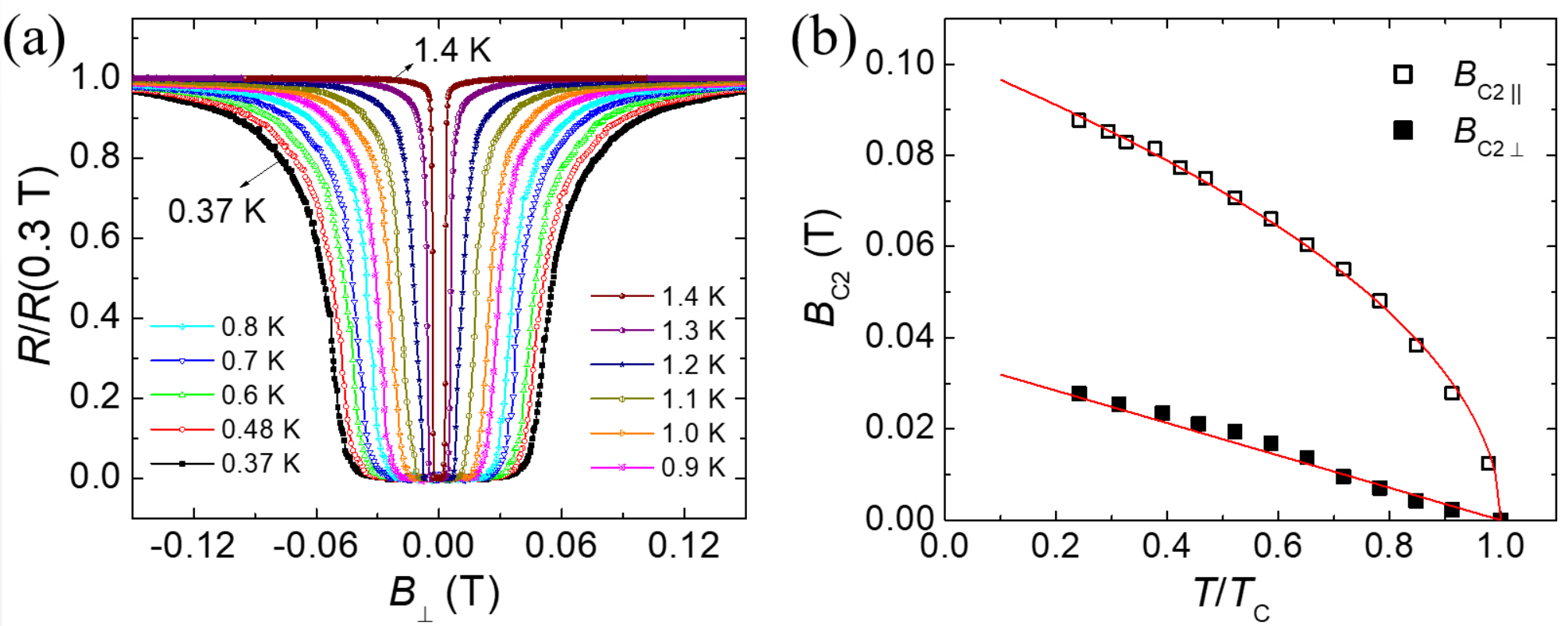} 
\caption{Upper critical fields of a 105-nm thick CoSi$_2$/Si(100) film.
(a) Normalized resistance $R(B)/R$(0.3\,T) versus perpendicular $B$ field at several $T$ values. (b) Perpendicular and parallel upper critical fields versus reduced temperature $T/T_c$ ($T_c$ = 1.534 K). The straight solid line and the solid curve through $B_{c2,\perp}$ and $B_{c2,\parallel}$, respectively, are least-squares fits (see text).
}
\label{fig10}
\end{figure}

We have measured the upper critical fields in perpendicular and parallel directions for a series of CoSi$_2$/Si films. Figure \ref{fig10}(a) shows the normalized resistance, $R(B)/R(B$\,=\,0.3\,T), as a function of magnetic field for a representative 105-nm thick CoSi$_2$/Si(100) film in perpendicular $B$ fields and at several $T$ values. We define the critical field as the field where the sample resistance deviates from zero. The variation of $B_{c2,\perp}$ with $T$ can be determined from this figure. Similarly, $B_{c2,\parallel}(T)$ can be measured in parallel $B$ fields.    

Figure \ref{fig10}(b) shows the perpendicular and parallel upper critical fields as a function of temperature for the film shown in Fig. \ref{fig10}(a). As expected for a superconducting film \cite{Cohn1988}, it is clearly seen that $B_{c2,\perp}$ obeys a linear $T$ dependence which can be expressed by $B_{c2,\perp}(T) = B_{c2,\perp}(0)(1-T/T_c)$, while $B_{c2,\parallel}$ obeys a square-root-$T$ dependence which can be expressed by $B_{c2,\parallel}(T) = B_{c2,\parallel}(0) (1-T/T_c)^{1/2}$. For this film, we obtain the extrapolated zero temperature fields $B_{c2,\perp}$(0) = 0.033 T and $B_{c2,\parallel}$(0) = 0.10 T. The upper critical fields of several films are listed in Table \ref{table2}.

\textbf{Ginzburg-Landau coherence length.} We have evaluated the superconducting coherence length $\xi_{\rm GL}(T)$ in the film plane from the perpendicular upper critical field through the Ginzburg-Landau relation 
$B_{c2,\perp}(T) = \phi_0/2\pi\xi_{\rm GL}^2(T)$, where $\phi_0 = h/2e$ is the flux quantum. Our extracted values for $\xi_{\rm GL}$ as $T$\,$\rightarrow$\,0 K are listed in Table \ref{table2}. Inspection of Table \ref{table2} indicates that $\xi_{\rm GL}(T$\,$\rightarrow$\,0) gradually decreases from 100 nm to 81 nm as the CoSi$_2$/Si film thickness decreases from 105 to 7 nm, insensitive to Si substrate orientation. For all films studied in this work, the extracted coherence length is longer than the elastic carrier mean free path, {\it i.e.}, $\xi_{\rm GL}(T$\,$\rightarrow$\,0) $> l_e$. Similarly, the superconducting coherence length in the direction normal to the film plane, $\xi_\perp (T)$, can be evaluated through the Ginzburg-Landau relation $B_{c2,\parallel} = \phi_0/(2\pi\xi_{\rm GL} \xi_\perp$), see Table \ref{table2}.

\textbf{Penetration depth.} The London penetration depth in the clean limit can be evaluated from the Ginzburg-Landau expression: $\lambda_{\rm L}(0) = \sqrt{ \tilde{m} /(\mu_0 \tilde{n}_s \tilde{e}^2})$, where $\tilde{m}$ ($\tilde{e}$) is the mass (charge) of a Cooper pair, $\tilde{n}_s$ is the density of Cooper pairs, and $\mu_0$ is the permeability of free space. We obtain $\lambda_{\rm L}(0) \simeq$ 38 nm, with $\tilde{n}_s \simeq 1 \times 10^{28}$ m$^{-3}$. When the elastic carrier mean free path has a finite value, the penetration depth is modified and the BCS theory predicts $\lambda (T$\,=\,0) = $\lambda_{\rm L}(0) \sqrt{1 + \xi_0/l_e}$, where $\xi_0 = \hbar v_F/\pi \triangle_0$ is the coherence length in the BCS theory \cite{Tinkham2004}. For example, for our 105- (24.5-) nm thick CoSi$_2$/Si film, we estimate a value $\lambda (0) \simeq$ 140 (220) nm, see Table \ref{table2}. These evaluations of $\lambda (0)$ may be subject to some uncertainties. Nevertheless, it should be safe to conclude that the CoSi$_2$/Si films are type II superconductors. They fall deeper in the type-II regime when the films are made thinner. Experimentally, values of $\lambda (T \rightarrow 0)$ may be inferred from the measurements of effective microwave surface impedance by employing a resonator technique \cite{Gubin2005} or measurements of the vortex inductance in a type II superconductor \cite{Fuchs2022}. These will be addressed in future work.

In a spin-singlet superconductor, the $B_{c2}(T$\,$\rightarrow$\,0) cannot exceed the Pauli or Clogston–Chandrasekhar paramagnetic limit, which in the BCS theory is given by $(B_{c2})_{\rm Pauli}$\,=\,1.84\,$T_c$, where $B_{c2}$ is in T and $T_c$ is in K \cite{Tinkham2004}. For a $p$-wave superconductor, the possible phases are distinguished by the spin vector $\vec{d}$. The value of $B_{c2}$ at zero temperature varies for each phase. The polar phase, which is formed by Cooper pairs having  $S_z=0$ (spin) and $m=0$ (orbital) angular momentum, has the highest $B_{c2}$ value. It surpasses the Pauli or Clogston–Chandrasekhar paramagnetic limit \cite{Scharnberg.80,Burlachkov.85}. For the two-component superconductivity, the breaking of inversion symmetry can lead to helical phases \cite{Mineev.94} and the orientation of $\vec{B}$ with respect to $\vec{\mathcal{A}}$ matters \cite{Mishra.21,Mishra.23}. Experiments exploring the anisotropy of $\vec{B}_{c2}(T)$ are currently ongoing.

\begin{widetext}
\begin{table*}
\caption{\label{table2}
Relevant parameters for five representative CoSi$_2$/Si films. $t$ is the film thickness, $T_c$ is the superconducting transition temperature, $\rho_0$ is the residual resistivity, $l_e$ is the elastic mean free path, $B_{c2\perp}$ ($B_{c2,\parallel}$) is the perpendicular (parallel) upper critical field, $\xi_{\rm GL}$ is the Ginzburg-Landau coherence length in the film plane, and $\lambda (0)$ is the penetration depth at $T$\,=\,0. The listed $B_{c2,\perp}$, $B_{c2,\parallel}$ and $\xi_{\rm GL}$ values were those extrapolated as $T$\,$\rightarrow$\,0. The $\lambda (0)$ values were calculated from the BCS theory (see text). 
}
\begin{ruledtabular}
\begin{tabular}{lcccccccc}
$t$ (nm) & Si & $T_c$\,(K) & $\rho_0$\,($\mu\Omega$ cm) & $l_e$\,(nm) & $B_{c2,\perp}$\,(T) & $B_{c2,\parallel}$\,(T) & $\xi_{\rm GL}$\,(nm) & $\lambda (0)$\,(nm) \\ 
\hline 
105 & (100) & 1.54 & 2.5 & 67 & 0.033 & 0.10 & 100 & 140\\
52.5 & (100) & 1.35 & 3.5 & 48 & 0.032 & -- & 100 & 170\\
35  & (111) & 1.18 & 4.3 & 39 & 0.040 & 0.22 & 91 & 200\\
24.5 & (100) & 1.25 & 5.5 & 30 & 0.044 & -- & 87 & 220\\
7   & (111) & 1.15 & 11 & 15 & 0.050 & 0.78 & 81 & 330\\
\end{tabular}
\end{ruledtabular}
\end{table*}
\end{widetext}

\subsection{Superconducting critical current density}

\begin{center}
\begin{figure}[t!]
\includegraphics[width=0.48\textwidth]{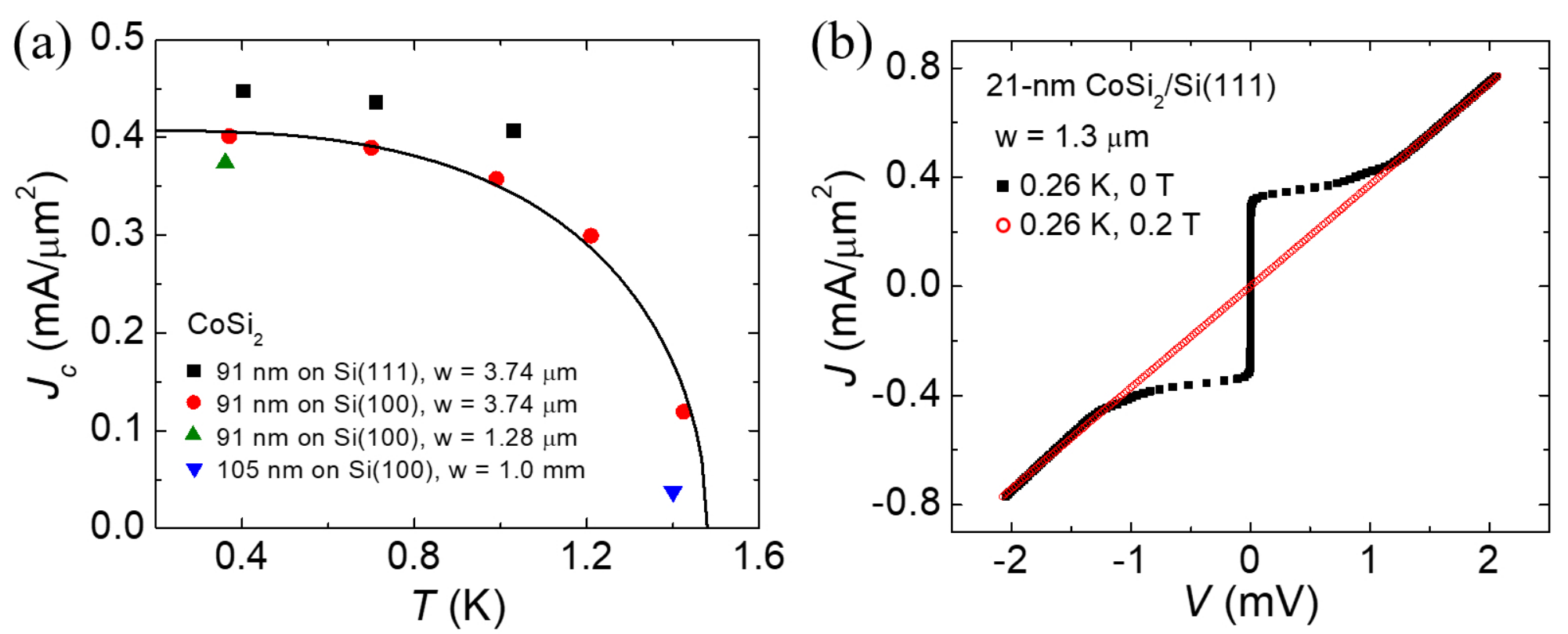} 
\caption{(a) Superconducting critical current density $J_c$ versus $T$ for four CoSi$_2$/Si films. The solid curve through the red circles is a least-squares fit to the 91-nm thick and 3.74-$\mu$m wide CoSi$_2$/Si(100) film. (b) $J$-$V$ curves for a CoSi$_2$/Si(111) microbridge measured in zero $B$ field and in a perpendicular field $B$ = 0.2 T.
}
\label{fig11}
\end{figure}  
\end{center}

We have measured the superconducting critical current density ($J_c$) for several CoSi$_2$/Si films in zero magnetic field. Figure \ref{fig11}(a) shows the variation of $J_c$ with $T$ for three 91-nm and one 105-nm thick films, as indicated. The solid curve through one of the samples (red circles) is the least-squares fit to the phenomenological expression: $J_{c}(T) = J_{c}(0)\, {\rm tanh}[b \sqrt{(T_c/T-1)}]$, where $b$ is a dimensionless parameter of the order of $\triangle_0/k_BT_c$, and $k_B$ is the Boltzmann constant \cite{Ruvalds1996}. Good agreement is obtained, with extrapolated $J_c(T$\,=\,0) = 0.41 mA/$\mu$m$^2$. We have also measured a 21-nm thick and 1.3-$\mu$m wide CoSi$_2$/Si(111) microbridge with $\rho_0$ = 5.4 $\mu\Omega$ cm, see Fig. \ref{fig11}(b). The measured $J_c$(0.26\,K) $\simeq$ 0.38 mA/$\mu$m$^2$ in $B$ = 0 (black symbols) is close to that shown in Fig. \ref{fig11}(a). In a perpendicular magnetic field $B$ = 0.2 T (red straight line), the microbridge returns to the normal state.

\subsection{Superconducting microwave resonators}

Superconducting microwave resonators have versatile and indispensable applications in qubit readout \cite{Wallraff2004}, qubit-qubit coupling \cite{Mariantoni2011}, quantum memory \cite{Hofheinz2009}, and sensitive photon detectors \cite{Day2003}, etc. These emergent applications have inspired intense efforts on the search for novel material candidates and fabrication techniques to implement high-performance superconducting devices. In practice, the performance of superconducting microwave resonators is deemed to be hampered by the loss and low frequency 1/$f$ noise, which are believed to arise from two-level systems. However, the detrimental two-level systems in superconducting circuits are yet to be fully identified and categorized, and then to be removed or their number minimized \cite{McRae2020}. In this regard, the appealing material properties of CoSi$_2$/Si films reported in this paper, especially the ultralow 1/$f$ noise amplitude, suggest their potential for reaching a high quality factor and small frequency noise in superconducting microwave resonators. Moreover, the spin-triplet chiral $p$-wave pairing symmetry observed in CoSi$_2$/TiSi$_2$ S/N junctions and T-shaped proximity structures aforementioned \cite{Chiu.21, Chiu2023} suggest that CoSi$_2$/Si films may be useful for hosting topological superconducting properties that are highly desired for the quantum computing technology \cite{Stern2013}. Experiments in this direction are encouraged.

\subsection{Arc-melted bulk CoSi$_2$}

\begin{center}
\begin{figure}[t!]
\includegraphics[width=0.45\textwidth]{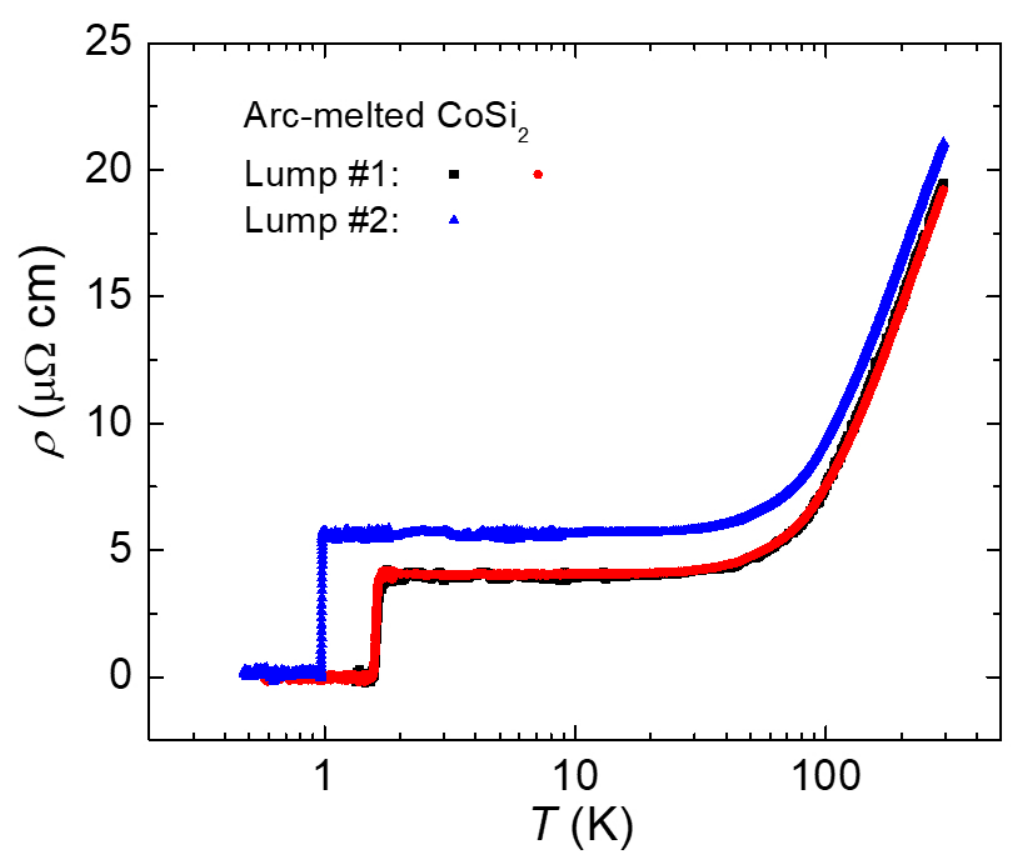} 
\caption{$\rho$ versus log($T$) for three polycrystalline CoSi$_2$ bulk samples prepared via the arc-melting method from two source lumps. 
}
\label{fig12}
\end{figure}  
\end{center}

For comparison, we have prepared several bulk samples by the standard arc-melting method \cite{Wu1994} from two sources of commercial CoSi$_2$ lumps. Figure \ref{fig12} shows the variation of resistivity with log($T$). All samples are metallic, undergoing superconducting at low temperatures. It is clear that $\rho(T)$ of all samples approaches $\rho_0$ at about 40 K, a $T$ value higher than the value where our CoSi$_2$/Si films reach $\rho_0$ [cf. Fig. \ref{fig3}(a)]. Interestingly, for the two bulk samples (red and black symbols) having a $T_c$ value as high as $\simeq$ 1.6 K, their $\rho_0$ value (= 4.0 $\mu\Omega$ cm) are larger than that (1.77 $\mu\Omega$ cm) of our thickest CoSi$_2$/Si films [cf. Fig. \ref{fig3}(c)]. This result indicates that our films are really of high quality and they are nominally epitaxial \cite{Chiu.17}. The third sample (blue symbols) in Fig. \ref{fig12} was arc-melted from a source lump different from that of the first two samples. This bulk sample has a relatively large $\rho_0$ value (close to that of a 10-nm thick CoSi$_2$/Si film) and a low $T_c \simeq$ 1.0 K (lower than that of all our films shown in Fig. \ref{fig9}). The reason for such a low $T_c$ value is due to the source lump having a low purity. The atomic emission spectroscopy studies indicate that the source lump contains magnetic and nonmagnetic impurities, including, among others, 60 ppm of Fe and 35 ppm of Ni. On the other hand, the first source lump reveals a X-ray diffraction pattern similar to that of Fig. \ref{fig2}(a), indicating that it is a relatively clean and single-phased CoSi$_2$. The $T_c$ values obtained from the two clean samples are also plotted in Fig. \ref{fig9} (blue down triangles).

\section{Conclusion}

Thin films of nominally epitaxial CoSi$_2$ grown on silicon Si(100) and Si(111) substrates reveal unusual normal-state electrical-transport characteristics. In addition to showing typical metallic conduction with low resistivities, they process unexpectedly strong spin-orbit coupling, manifesting notable positive magnetoresistance arising from the weak-antilocalization effect. They have a long electron dephasing length exceeding 1 $\mu$m at 4 K. They show ultralow 1/$f$ noise, which can be attributed to the strong covalent bonding in the CoSi$_2$ crystal structure as well as a small lattice mismatch between CoSi$_2$ and silicon. The strong spin-orbit coupling is likely of the Rashba type, which induces two-component ($s$ + $p$) superconductivity with the spin-triplet $p$-wave pairing component being dominant, as previously established from the conductance spectroscopy studies of CoSi$_2$/TiSi$_2$ S/N heterojunctions \cite{Chiu.21,Mishra.21} and T-shaped superconducting proximity structures \cite{Chiu.21,Chiu2023}. The long electron mean free path and ultralow 1/$f$ noise shall benefit the potential applications of CoSi$_2$/Si films as the building blocks of superconducting and quantum devices. In contrast to other material candidates of unconventional (topological) superconductors, which are usually bulk crystals, the growth of CoSi$_2$/Si films are fully compatible with the present-day silicon-based integrated-circuit technology. In particular, the stable thin-film form in the ambient allows them for micro-fabrication and patterning via the standard photo-lithographic and electron-beam lithographic technique, promising device scalability in superconducting circuits. We would like to propose that (nominally) epitaxial CoSi$_2$/Si/CoSi$_2$ Josephson junctions and CoSi$_2$/CoSi/CoSi$_2$ Josephson weak links be developed and explored, where CoSi is a topological semimetal \cite{Tang2017,Pshenay2018}. Superconducting devices based on CoSi$_2$/Si films may open an avenue for fundamental research and applications in the emergent quantum-information  technology. \\




\begin{acknowledgments}

\begin{center}
    \textbf{ACKNOWLEDGMENTS}
\end{center}

We thank C.\,H.,Hsu and C.\,H.\,Lin for the measurement of Fig. \ref{fig11}(b), C. W. Wu for fabricating samples for the 1/$f$ noise measurement, and  Y.\,Asano, P.\,Hakonen, V.\,Mishra, C.\,Strunk, Y.\,Tanaka, and F.\,C.\,Zhang for discussion. This work was supported by the National Science and Technology Council of Taiwan through grant numbers 110-2112-M-A49-015 and 111-2119-M-007-005 (J.J.L.), 110-2112-M-A49-033-MY3 (S.S.Y.), and 112-2112-M-A49-MY4  (S.K.). S.K. acknowledges support by the Yushan Fellowship Program of the Ministry of Education (MOE) of Taiwan and the Center for Theoretical and Computational Physics of NYCU, Taiwan. S.S.Y. was partly supported by the Center for Emergent Functional Matter Science of NYCU from The Featured Areas Research Center Program within the framework of the Higher Education Sprout Project by the MOE.

\end{acknowledgments}



%


\begin{thebibliography}{75}%
\makeatletter
\providecommand \@ifxundefined [1]{%
 \@ifx{#1\undefined}
}%
\providecommand \@ifnum [1]{%
 \ifnum #1\expandafter \@firstoftwo
 \else \expandafter \@secondoftwo
 \fi
}%
\providecommand \@ifx [1]{%
 \ifx #1\expandafter \@firstoftwo
 \else \expandafter \@secondoftwo
 \fi
}%
\providecommand \natexlab [1]{#1}%
\providecommand \enquote  [1]{``#1''}%
\providecommand \bibnamefont  [1]{#1}%
\providecommand \bibfnamefont [1]{#1}%
\providecommand \citenamefont [1]{#1}%
\providecommand \href@noop [0]{\@secondoftwo}%
\providecommand \href [0]{\begingroup \@sanitize@url \@href}%
\providecommand \@href[1]{\@@startlink{#1}\@@href}%
\providecommand \@@href[1]{\endgroup#1\@@endlink}%
\providecommand \@sanitize@url [0]{\catcode `\\12\catcode `\$12\catcode `\&12\catcode `\#12\catcode `\^12\catcode `\_12\catcode `\%12\relax}%
\providecommand \@@startlink[1]{}%
\providecommand \@@endlink[0]{}%
\providecommand \url  [0]{\begingroup\@sanitize@url \@url }%
\providecommand \@url [1]{\endgroup\@href {#1}{\urlprefix }}%
\providecommand \urlprefix  [0]{URL }%
\providecommand \Eprint [0]{\href }%
\providecommand \doibase [0]{http://dx.doi.org/}%
\providecommand \selectlanguage [0]{\@gobble}%
\providecommand \bibinfo  [0]{\@secondoftwo}%
\providecommand \bibfield  [0]{\@secondoftwo}%
\providecommand \translation [1]{[#1]}%
\providecommand \BibitemOpen [0]{}%
\providecommand \bibitemStop [0]{}%
\providecommand \bibitemNoStop [0]{.\EOS\space}%
\providecommand \EOS [0]{\spacefactor3000\relax}%
\providecommand \BibitemShut  [1]{\csname bibitem#1\endcsname}%
\let\auto@bib@innerbib\@empty
\bibitem [{\citenamefont {Qi}\ and\ \citenamefont {Zhang}(2011)}]{Qi2011}%
  \BibitemOpen
  \bibfield  {author} {\bibinfo {author} {\bibfnamefont {X.-L.}\ \bibnamefont {Qi}}\ and\ \bibinfo {author} {\bibfnamefont {S.-C.}\ \bibnamefont {Zhang}},\ }\href {\doibase 10.1103/RevModPhys.83.1057} {\bibfield  {journal} {\bibinfo  {journal} {Rev. Mod. Phys.}\ }\textbf {\bibinfo {volume} {83}},\ \bibinfo {pages} {1057} (\bibinfo {year} {2011})}\BibitemShut {NoStop}%
\bibitem [{\citenamefont {Kallin}\ and\ \citenamefont {Berlinsky}(2016)}]{Kallin2016}%
  \BibitemOpen
  \bibfield  {author} {\bibinfo {author} {\bibfnamefont {C.}~\bibnamefont {Kallin}}\ and\ \bibinfo {author} {\bibfnamefont {J.}~\bibnamefont {Berlinsky}},\ }\href {\doibase 10.1088/0034-4885/79/5/054502} {\bibfield  {journal} {\bibinfo  {journal} {Rep. Prog. Phys.}\ }\textbf {\bibinfo {volume} {79}},\ \bibinfo {pages} {054502} (\bibinfo {year} {2016})}\BibitemShut {NoStop}%
\bibitem [{\citenamefont {Stewart}(2017)}]{Stewart2017}%
  \BibitemOpen
  \bibfield  {author} {\bibinfo {author} {\bibfnamefont {G.~R.}\ \bibnamefont {Stewart}},\ }\href {\doibase 10.1080/00018732.2017.1331615} {\bibfield  {journal} {\bibinfo  {journal} {Adv. Phys.}\ }\textbf {\bibinfo {volume} {66}},\ \bibinfo {pages} {75} (\bibinfo {year} {2017})}\BibitemShut {NoStop}%
\bibitem [{\citenamefont {Linder}\ and\ \citenamefont {Balatsky}(2019)}]{Linder2019}%
  \BibitemOpen
  \bibfield  {author} {\bibinfo {author} {\bibfnamefont {J.}~\bibnamefont {Linder}}\ and\ \bibinfo {author} {\bibfnamefont {A.~V.}\ \bibnamefont {Balatsky}},\ }\href {\doibase 10.1103/RevModPhys.91.045005} {\bibfield  {journal} {\bibinfo  {journal} {Rev. Mod. Phys.}\ }\textbf {\bibinfo {volume} {91}},\ \bibinfo {pages} {045005} (\bibinfo {year} {2019})}\BibitemShut {NoStop}%
\bibitem [{\citenamefont {Tanaka}\ and\ \citenamefont {Kashiwaya}(2004)}]{Tanaka2004anomalous}%
  \BibitemOpen
  \bibfield  {author} {\bibinfo {author} {\bibfnamefont {Y.}~\bibnamefont {Tanaka}}\ and\ \bibinfo {author} {\bibfnamefont {S.}~\bibnamefont {Kashiwaya}},\ }\href {\doibase 10.1103/PhysRevB.70.012507} {\bibfield  {journal} {\bibinfo  {journal} {Phys. Rev. B}\ }\textbf {\bibinfo {volume} {70}},\ \bibinfo {pages} {012507} (\bibinfo {year} {2004})}\BibitemShut {NoStop}%
\bibitem [{\citenamefont {Tanaka}\ \emph {et~al.}(2005)\citenamefont {Tanaka}, \citenamefont {Kashiwaya},\ and\ \citenamefont {Yokoyama}}]{Tanaka2005theory}%
  \BibitemOpen
  \bibfield  {author} {\bibinfo {author} {\bibfnamefont {Y.}~\bibnamefont {Tanaka}}, \bibinfo {author} {\bibfnamefont {S.}~\bibnamefont {Kashiwaya}}, \ and\ \bibinfo {author} {\bibfnamefont {T.}~\bibnamefont {Yokoyama}},\ }\href {\doibase 10.1103/PhysRevB.71.094513} {\bibfield  {journal} {\bibinfo  {journal} {Phys. Rev. B}\ }\textbf {\bibinfo {volume} {71}},\ \bibinfo {pages} {094513} (\bibinfo {year} {2005})}\BibitemShut {NoStop}%
\bibitem [{\citenamefont {Chiu}\ \emph {et~al.}(2021{\natexlab{a}})\citenamefont {Chiu}, \citenamefont {Tsuei}, \citenamefont {Yeh}, \citenamefont {Zhang}, \citenamefont {Kirchner},\ and\ \citenamefont {Lin}}]{Chiu.21}%
  \BibitemOpen
  \bibfield  {author} {\bibinfo {author} {\bibfnamefont {S.-P.}\ \bibnamefont {Chiu}}, \bibinfo {author} {\bibfnamefont {C.~C.}\ \bibnamefont {Tsuei}}, \bibinfo {author} {\bibfnamefont {S.-S.}\ \bibnamefont {Yeh}}, \bibinfo {author} {\bibfnamefont {F.-C.}\ \bibnamefont {Zhang}}, \bibinfo {author} {\bibfnamefont {S.}~\bibnamefont {Kirchner}}, \ and\ \bibinfo {author} {\bibfnamefont {J.-J.}\ \bibnamefont {Lin}},\ }\href {\doibase 10.1126/sciadv.abg6569} {\bibfield  {journal} {\bibinfo  {journal} {Sci. Adv.}\ }\textbf {\bibinfo {volume} {7}},\ \bibinfo {pages} {eabg6569} (\bibinfo {year} {2021}{\natexlab{a}})}\BibitemShut {NoStop}%
\bibitem [{\citenamefont {Chiu}\ \emph {et~al.}(2023)\citenamefont {Chiu}, \citenamefont {Mishra}, \citenamefont {Li}, \citenamefont {Zhang}, \citenamefont {Kirchner},\ and\ \citenamefont {Lin}}]{Chiu2023}%
  \BibitemOpen
  \bibfield  {author} {\bibinfo {author} {\bibfnamefont {S.-P.}\ \bibnamefont {Chiu}}, \bibinfo {author} {\bibfnamefont {V.}~\bibnamefont {Mishra}}, \bibinfo {author} {\bibfnamefont {Y.}~\bibnamefont {Li}}, \bibinfo {author} {\bibfnamefont {F.-C.}\ \bibnamefont {Zhang}}, \bibinfo {author} {\bibfnamefont {S.}~\bibnamefont {Kirchner}}, \ and\ \bibinfo {author} {\bibfnamefont {J.-J.}\ \bibnamefont {Lin}},\ }\href {\doibase 10.1039/D2NR05864B} {\bibfield  {journal} {\bibinfo  {journal} {Nanoscale}\ }\textbf {\bibinfo {volume} {15}},\ \bibinfo {pages} {9179} (\bibinfo {year} {2023})}\BibitemShut {NoStop}%
\bibitem [{\citenamefont {Chiu}\ \emph {et~al.}(2021{\natexlab{b}})\citenamefont {Chiu}, \citenamefont {Lai},\ and\ \citenamefont {Lin}}]{Chiu2021b}%
  \BibitemOpen
  \bibfield  {author} {\bibinfo {author} {\bibfnamefont {S.-P.}\ \bibnamefont {Chiu}}, \bibinfo {author} {\bibfnamefont {W.-L.}\ \bibnamefont {Lai}}, \ and\ \bibinfo {author} {\bibfnamefont {J.-J.}\ \bibnamefont {Lin}},\ }\href {\doibase 10.35848/1347-4065/ac1693} {\bibfield  {journal} {\bibinfo  {journal} {Jpn. J. Appl. Phys.}\ }\textbf {\bibinfo {volume} {60}},\ \bibinfo {pages} {088002} (\bibinfo {year} {2021}{\natexlab{b}})}\BibitemShut {NoStop}%
\bibitem [{\citenamefont {Tsutsumi}\ \emph {et~al.}(1995)\citenamefont {Tsutsumi}, \citenamefont {Takayanagi}, \citenamefont {Ishikawa},\ and\ \citenamefont {Hirano}}]{tsutsumi1995}%
  \BibitemOpen
  \bibfield  {author} {\bibinfo {author} {\bibfnamefont {K.}~\bibnamefont {Tsutsumi}}, \bibinfo {author} {\bibfnamefont {S.}~\bibnamefont {Takayanagi}}, \bibinfo {author} {\bibfnamefont {M.}~\bibnamefont {Ishikawa}}, \ and\ \bibinfo {author} {\bibfnamefont {T.}~\bibnamefont {Hirano}},\ }\href {\doibase 10.1143/JPSJ.64.2237} {\bibfield  {journal} {\bibinfo  {journal} {J. Phys. Soc. Japan}\ }\textbf {\bibinfo {volume} {64}},\ \bibinfo {pages} {2237} (\bibinfo {year} {1995})}\BibitemShut {NoStop}%
\bibitem [{\citenamefont {Allen}\ and\ \citenamefont {Schulz}(1993)}]{Allen1993}%
  \BibitemOpen
  \bibfield  {author} {\bibinfo {author} {\bibfnamefont {P.~B.}\ \bibnamefont {Allen}}\ and\ \bibinfo {author} {\bibfnamefont {W.~W.}\ \bibnamefont {Schulz}},\ }\href {\doibase 10.1103/PhysRevB.47.14434} {\bibfield  {journal} {\bibinfo  {journal} {Phys. Rev. B}\ }\textbf {\bibinfo {volume} {47}},\ \bibinfo {pages} {14434} (\bibinfo {year} {1993})}\BibitemShut {NoStop}%
\bibitem [{\citenamefont {Asano}\ \emph {et~al.}(2007)\citenamefont {Asano}, \citenamefont {Tanaka}, \citenamefont {Golubov},\ and\ \citenamefont {Kashiwaya}}]{Asano2007}%
  \BibitemOpen
  \bibfield  {author} {\bibinfo {author} {\bibfnamefont {Y.}~\bibnamefont {Asano}}, \bibinfo {author} {\bibfnamefont {Y.}~\bibnamefont {Tanaka}}, \bibinfo {author} {\bibfnamefont {A.~A.}\ \bibnamefont {Golubov}}, \ and\ \bibinfo {author} {\bibfnamefont {S.}~\bibnamefont {Kashiwaya}},\ }\href {\doibase 10.1103/PhysRevLett.99.067005} {\bibfield  {journal} {\bibinfo  {journal} {Phys. Rev. Lett.}\ }\textbf {\bibinfo {volume} {99}},\ \bibinfo {pages} {067005} (\bibinfo {year} {2007})}\BibitemShut {NoStop}%
\bibitem [{\citenamefont {Tanaka}\ \emph {et~al.}(2007)\citenamefont {Tanaka}, \citenamefont {Tanuma},\ and\ \citenamefont {Golubov}}]{Tanaka2007}%
  \BibitemOpen
  \bibfield  {author} {\bibinfo {author} {\bibfnamefont {Y.}~\bibnamefont {Tanaka}}, \bibinfo {author} {\bibfnamefont {Y.}~\bibnamefont {Tanuma}}, \ and\ \bibinfo {author} {\bibfnamefont {A.~A.}\ \bibnamefont {Golubov}},\ }\href {\doibase 10.1103/PhysRevB.76.054522} {\bibfield  {journal} {\bibinfo  {journal} {Phys. Rev. B}\ }\textbf {\bibinfo {volume} {76}},\ \bibinfo {pages} {054522} (\bibinfo {year} {2007})}\BibitemShut {NoStop}%
\bibitem [{\citenamefont {Blonder}\ \emph {et~al.}(1982)\citenamefont {Blonder}, \citenamefont {Tinkham},\ and\ \citenamefont {Klapwijk}}]{BTK1982}%
  \BibitemOpen
  \bibfield  {author} {\bibinfo {author} {\bibfnamefont {G.~E.}\ \bibnamefont {Blonder}}, \bibinfo {author} {\bibfnamefont {M.}~\bibnamefont {Tinkham}}, \ and\ \bibinfo {author} {\bibfnamefont {T.~M.}\ \bibnamefont {Klapwijk}},\ }\href {\doibase 10.1103/PhysRevB.25.4515} {\bibfield  {journal} {\bibinfo  {journal} {Phys. Rev. B}\ }\textbf {\bibinfo {volume} {25}},\ \bibinfo {pages} {4515} (\bibinfo {year} {1982})}\BibitemShut {NoStop}%
\bibitem [{\citenamefont {Yamashiro}\ \emph {et~al.}(1997)\citenamefont {Yamashiro}, \citenamefont {Tanaka},\ and\ \citenamefont {Kashiwaya}}]{Yamashiro1997}%
  \BibitemOpen
  \bibfield  {author} {\bibinfo {author} {\bibfnamefont {M.}~\bibnamefont {Yamashiro}}, \bibinfo {author} {\bibfnamefont {Y.}~\bibnamefont {Tanaka}}, \ and\ \bibinfo {author} {\bibfnamefont {S.}~\bibnamefont {Kashiwaya}},\ }\href {\doibase 10.1103/PhysRevB.56.7847} {\bibfield  {journal} {\bibinfo  {journal} {Phys. Rev. B}\ }\textbf {\bibinfo {volume} {56}},\ \bibinfo {pages} {7847} (\bibinfo {year} {1997})}\BibitemShut {NoStop}%
\bibitem [{\citenamefont {Courtois}\ \emph {et~al.}(1999)\citenamefont {Courtois}, \citenamefont {Charlat}, \citenamefont {Gandit}, \citenamefont {Mailly},\ and\ \citenamefont {Pannetier}}]{Courtois1999}%
  \BibitemOpen
  \bibfield  {author} {\bibinfo {author} {\bibfnamefont {H.}~\bibnamefont {Courtois}}, \bibinfo {author} {\bibfnamefont {P.}~\bibnamefont {Charlat}}, \bibinfo {author} {\bibfnamefont {P.}~\bibnamefont {Gandit}}, \bibinfo {author} {\bibfnamefont {D.}~\bibnamefont {Mailly}}, \ and\ \bibinfo {author} {\bibfnamefont {B.}~\bibnamefont {Pannetier}},\ }\href {\doibase 10.1023/A:1021885617107} {\bibfield  {journal} {\bibinfo  {journal} {J. Low Temp. Phys.}\ }\textbf {\bibinfo {volume} {116}},\ \bibinfo {pages} {187} (\bibinfo {year} {1999})}\BibitemShut {NoStop}%
\bibitem [{\citenamefont {Vollhardt}\ and\ \citenamefont {Woelfle}(1990)}]{VollhardtWoefle}%
  \BibitemOpen
  \bibfield  {author} {\bibinfo {author} {\bibfnamefont {D.}~\bibnamefont {Vollhardt}}\ and\ \bibinfo {author} {\bibfnamefont {P.}~\bibnamefont {Woelfle}},\ }\href {\doibase https://doi.org/10.1201/b12808} {\emph {\bibinfo {title} {The Superfluid Phases Of Helium 3}}}\ (\bibinfo  {publisher} {CRC Press},\ \bibinfo {year} {1990})\BibitemShut {NoStop}%
\bibitem [{\citenamefont {Shimizu}\ \emph {et~al.}(2019)\citenamefont {Shimizu}, \citenamefont {Braithwaite}, \citenamefont {Aoki}, \citenamefont {Salce},\ and\ \citenamefont {Brison}}]{Shimizu2019}%
  \BibitemOpen
  \bibfield  {author} {\bibinfo {author} {\bibfnamefont {Y.}~\bibnamefont {Shimizu}}, \bibinfo {author} {\bibfnamefont {D.}~\bibnamefont {Braithwaite}}, \bibinfo {author} {\bibfnamefont {D.}~\bibnamefont {Aoki}}, \bibinfo {author} {\bibfnamefont {B.}~\bibnamefont {Salce}}, \ and\ \bibinfo {author} {\bibfnamefont {J.-P.}\ \bibnamefont {Brison}},\ }\href {\doibase 10.1103/PhysRevLett.122.067001} {\bibfield  {journal} {\bibinfo  {journal} {Phys. Rev. Lett.}\ }\textbf {\bibinfo {volume} {122}},\ \bibinfo {pages} {067001} (\bibinfo {year} {2019})}\BibitemShut {NoStop}%
\bibitem [{\citenamefont {Schemm}\ \emph {et~al.}(2014)\citenamefont {Schemm}, \citenamefont {Gannon}, \citenamefont {Wishne}, \citenamefont {Halperin},\ and\ \citenamefont {Kapitulnik}}]{Schemm.14}%
  \BibitemOpen
  \bibfield  {author} {\bibinfo {author} {\bibfnamefont {E.~R.}\ \bibnamefont {Schemm}}, \bibinfo {author} {\bibfnamefont {W.~J.}\ \bibnamefont {Gannon}}, \bibinfo {author} {\bibfnamefont {C.~M.}\ \bibnamefont {Wishne}}, \bibinfo {author} {\bibfnamefont {W.~P.}\ \bibnamefont {Halperin}}, \ and\ \bibinfo {author} {\bibfnamefont {A.}~\bibnamefont {Kapitulnik}},\ }\href {\doibase 10.1126/science.1248552} {\bibfield  {journal} {\bibinfo  {journal} {Science}\ }\textbf {\bibinfo {volume} {345}},\ \bibinfo {pages} {190} (\bibinfo {year} {2014})},\ \Eprint {http://arxiv.org/abs/https://www.science.org/doi/pdf/10.1126/science.1248552} {https://www.science.org/doi/pdf/10.1126/science.1248552} \BibitemShut {NoStop}%
\bibitem [{\citenamefont {Frigeri}\ \emph {et~al.}(2004)\citenamefont {Frigeri}, \citenamefont {Agterberg}, \citenamefont {Koga},\ and\ \citenamefont {Sigrist}}]{Frigeri2004}%
  \BibitemOpen
  \bibfield  {author} {\bibinfo {author} {\bibfnamefont {P.~A.}\ \bibnamefont {Frigeri}}, \bibinfo {author} {\bibfnamefont {D.~F.}\ \bibnamefont {Agterberg}}, \bibinfo {author} {\bibfnamefont {A.}~\bibnamefont {Koga}}, \ and\ \bibinfo {author} {\bibfnamefont {M.}~\bibnamefont {Sigrist}},\ }\href {\doibase 10.1103/PhysRevLett.92.097001} {\bibfield  {journal} {\bibinfo  {journal} {Phys. Rev. Lett.}\ }\textbf {\bibinfo {volume} {92}},\ \bibinfo {pages} {097001} (\bibinfo {year} {2004})}\BibitemShut {NoStop}%
\bibitem [{\citenamefont {Bauer}\ \emph {et~al.}(2004)\citenamefont {Bauer}, \citenamefont {Hilscher}, \citenamefont {Michor}, \citenamefont {Paul}, \citenamefont {Scheidt}, \citenamefont {Gribanov}, \citenamefont {Seropegin}, \citenamefont {No\"el}, \citenamefont {Sigrist},\ and\ \citenamefont {Rogl}}]{Bauer2004}%
  \BibitemOpen
  \bibfield  {author} {\bibinfo {author} {\bibfnamefont {E.}~\bibnamefont {Bauer}}, \bibinfo {author} {\bibfnamefont {G.}~\bibnamefont {Hilscher}}, \bibinfo {author} {\bibfnamefont {H.}~\bibnamefont {Michor}}, \bibinfo {author} {\bibfnamefont {C.}~\bibnamefont {Paul}}, \bibinfo {author} {\bibfnamefont {E.~W.}\ \bibnamefont {Scheidt}}, \bibinfo {author} {\bibfnamefont {A.}~\bibnamefont {Gribanov}}, \bibinfo {author} {\bibfnamefont {Y.}~\bibnamefont {Seropegin}}, \bibinfo {author} {\bibfnamefont {H.}~\bibnamefont {No\"el}}, \bibinfo {author} {\bibfnamefont {M.}~\bibnamefont {Sigrist}}, \ and\ \bibinfo {author} {\bibfnamefont {P.}~\bibnamefont {Rogl}},\ }\href {\doibase 10.1103/PhysRevLett.92.027003} {\bibfield  {journal} {\bibinfo  {journal} {Phys. Rev. Lett.}\ }\textbf {\bibinfo {volume} {92}},\ \bibinfo {pages} {027003} (\bibinfo {year} {2004})}\BibitemShut {NoStop}%
\bibitem [{\citenamefont {Silber}\ \emph {et~al.}(2024)\citenamefont {Silber}, \citenamefont {Mathimalar}, \citenamefont {Mangel}, \citenamefont {Nayak}, \citenamefont {Green}, \citenamefont {Avraham}, \citenamefont {Beidenkopf}, \citenamefont {Feldman}, \citenamefont {Kanigel}, \citenamefont {Klein}, \citenamefont {Goldstein}, \citenamefont {Banerjee}, \citenamefont {Sela},\ and\ \citenamefont {Dagan}}]{Silber.24}%
  \BibitemOpen
  \bibfield  {author} {\bibinfo {author} {\bibfnamefont {I.}~\bibnamefont {Silber}}, \bibinfo {author} {\bibfnamefont {S.}~\bibnamefont {Mathimalar}}, \bibinfo {author} {\bibfnamefont {I.}~\bibnamefont {Mangel}}, \bibinfo {author} {\bibfnamefont {A.~K.}\ \bibnamefont {Nayak}}, \bibinfo {author} {\bibfnamefont {O.}~\bibnamefont {Green}}, \bibinfo {author} {\bibfnamefont {N.}~\bibnamefont {Avraham}}, \bibinfo {author} {\bibfnamefont {H.}~\bibnamefont {Beidenkopf}}, \bibinfo {author} {\bibfnamefont {I.}~\bibnamefont {Feldman}}, \bibinfo {author} {\bibfnamefont {A.}~\bibnamefont {Kanigel}}, \bibinfo {author} {\bibfnamefont {A.}~\bibnamefont {Klein}}, \bibinfo {author} {\bibfnamefont {M.}~\bibnamefont {Goldstein}}, \bibinfo {author} {\bibfnamefont {A.}~\bibnamefont {Banerjee}}, \bibinfo {author} {\bibfnamefont {E.}~\bibnamefont {Sela}}, \ and\ \bibinfo {author} {\bibfnamefont {Y.}~\bibnamefont {Dagan}},\ }\href@noop {} {\bibfield  {journal} {\bibinfo  {journal} {Nature Communications}\ }\textbf {\bibinfo {volume}
  {15}},\ \bibinfo {pages} {824} (\bibinfo {year} {2024})}\BibitemShut {NoStop}%
\bibitem [{\citenamefont {Ommen}\ \emph {et~al.}(1988)\citenamefont {Ommen}, \citenamefont {Bulle‐Lieuwma},\ and\ \citenamefont {Langereis}}]{Ommen1988}%
  \BibitemOpen
  \bibfield  {author} {\bibinfo {author} {\bibfnamefont {A.~H.~v.}\ \bibnamefont {Ommen}}, \bibinfo {author} {\bibfnamefont {C.~W.~T.}\ \bibnamefont {Bulle‐Lieuwma}}, \ and\ \bibinfo {author} {\bibfnamefont {C.}~\bibnamefont {Langereis}},\ }\href {\doibase 10.1063/1.341612} {\bibfield  {journal} {\bibinfo  {journal} {J. Appl. Phys.}\ }\textbf {\bibinfo {volume} {64}},\ \bibinfo {pages} {2706} (\bibinfo {year} {1988})}\BibitemShut {NoStop}%
\bibitem [{\citenamefont {Chen}(2004)}]{Chen2004}%
  \BibitemOpen
  \bibfield  {author} {\bibinfo {author} {\bibfnamefont {L.~J.}\ \bibnamefont {Chen}},\ }\href@noop {} {\emph {\bibinfo {title} {Silicide technology for integrated circuits}}},\ Vol.~\bibinfo {volume} {5}\ (\bibinfo  {publisher} {Iet},\ \bibinfo {year} {2004})\BibitemShut {NoStop}%
\bibitem [{\citenamefont {Chen}\ and\ \citenamefont {Tu}(1991)}]{Chen1991}%
  \BibitemOpen
  \bibfield  {author} {\bibinfo {author} {\bibfnamefont {L.~J.}\ \bibnamefont {Chen}}\ and\ \bibinfo {author} {\bibfnamefont {K.-N.}\ \bibnamefont {Tu}},\ }\href {\doibase https://doi.org/10.1016/0920-2307(91)90004-7} {\bibfield  {journal} {\bibinfo  {journal} {Mat. Sci. Rep.}\ }\textbf {\bibinfo {volume} {6}},\ \bibinfo {pages} {53} (\bibinfo {year} {1991})}\BibitemShut {NoStop}%
\bibitem [{\citenamefont {Bulle‐Lieuwma}\ \emph {et~al.}(1992)\citenamefont {Bulle‐Lieuwma}, \citenamefont {van Ommen}, \citenamefont {Hornstra},\ and\ \citenamefont {Aussems}}]{Bulle1992}%
  \BibitemOpen
  \bibfield  {author} {\bibinfo {author} {\bibfnamefont {C.~W.~T.}\ \bibnamefont {Bulle‐Lieuwma}}, \bibinfo {author} {\bibfnamefont {A.~H.}\ \bibnamefont {van Ommen}}, \bibinfo {author} {\bibfnamefont {J.}~\bibnamefont {Hornstra}}, \ and\ \bibinfo {author} {\bibfnamefont {C.~N. A.~M.}\ \bibnamefont {Aussems}},\ }\href {\doibase 10.1063/1.351119} {\bibfield  {journal} {\bibinfo  {journal} {J. Appl. Phys.}\ }\textbf {\bibinfo {volume} {71}},\ \bibinfo {pages} {2211} (\bibinfo {year} {1992})}\BibitemShut {NoStop}%
\bibitem [{\citenamefont {Chiu}\ \emph {et~al.}(2017)\citenamefont {Chiu}, \citenamefont {Yeh}, \citenamefont {Chiou}, \citenamefont {Chou}, \citenamefont {Lin},\ and\ \citenamefont {Tsuei}}]{Chiu.17}%
  \BibitemOpen
  \bibfield  {author} {\bibinfo {author} {\bibfnamefont {S.-P.}\ \bibnamefont {Chiu}}, \bibinfo {author} {\bibfnamefont {S.-S.}\ \bibnamefont {Yeh}}, \bibinfo {author} {\bibfnamefont {C.-J.}\ \bibnamefont {Chiou}}, \bibinfo {author} {\bibfnamefont {Y.-C.}\ \bibnamefont {Chou}}, \bibinfo {author} {\bibfnamefont {J.-J.}\ \bibnamefont {Lin}}, \ and\ \bibinfo {author} {\bibfnamefont {C.-C.}\ \bibnamefont {Tsuei}},\ }\href {\doibase 10.1021/acsnano.6b06553} {\bibfield  {journal} {\bibinfo  {journal} {ACS Nano}\ }\textbf {\bibinfo {volume} {11}},\ \bibinfo {pages} {516} (\bibinfo {year} {2017})}\BibitemShut {NoStop}%
\bibitem [{\citenamefont {Lambrecht}\ \emph {et~al.}(1987)\citenamefont {Lambrecht}, \citenamefont {Christensen},\ and\ \citenamefont {Bl\"ochl}}]{Lambrecht1987}%
  \BibitemOpen
  \bibfield  {author} {\bibinfo {author} {\bibfnamefont {W.~R.~L.}\ \bibnamefont {Lambrecht}}, \bibinfo {author} {\bibfnamefont {N.~E.}\ \bibnamefont {Christensen}}, \ and\ \bibinfo {author} {\bibfnamefont {P.}~\bibnamefont {Bl\"ochl}},\ }\href {\doibase 10.1103/PhysRevB.36.2493} {\bibfield  {journal} {\bibinfo  {journal} {Phys. Rev. B}\ }\textbf {\bibinfo {volume} {36}},\ \bibinfo {pages} {2493} (\bibinfo {year} {1987})}\BibitemShut {NoStop}%
\bibitem [{\citenamefont {eon}\ \emph {et~al.}(1992)\citenamefont {eon}, \citenamefont {Sukow}, \citenamefont {Honeycutt}, \citenamefont {Rozgonyi},\ and\ \citenamefont {Nemanich}}]{Jeon1992}%
  \BibitemOpen
  \bibfield  {author} {\bibinfo {author} {\bibfnamefont {H.}~\bibnamefont {eon}}, \bibinfo {author} {\bibfnamefont {C.~A.}\ \bibnamefont {Sukow}}, \bibinfo {author} {\bibfnamefont {J.~W.}\ \bibnamefont {Honeycutt}}, \bibinfo {author} {\bibfnamefont {G.~A.}\ \bibnamefont {Rozgonyi}}, \ and\ \bibinfo {author} {\bibfnamefont {R.~J.}\ \bibnamefont {Nemanich}},\ }\href {\doibase 10.1063/1.350808} {\bibfield  {journal} {\bibinfo  {journal} {J. Appl. Phys.}\ }\textbf {\bibinfo {volume} {71}},\ \bibinfo {pages} {4269} (\bibinfo {year} {1992})}\BibitemShut {NoStop}%
\bibitem [{\citenamefont {Mammoliti}\ \emph {et~al.}(2002)\citenamefont {Mammoliti}, \citenamefont {Grimaldi},\ and\ \citenamefont {La~Via}}]{Mammoliti2002}%
  \BibitemOpen
  \bibfield  {author} {\bibinfo {author} {\bibfnamefont {F.}~\bibnamefont {Mammoliti}}, \bibinfo {author} {\bibfnamefont {M.~G.}\ \bibnamefont {Grimaldi}}, \ and\ \bibinfo {author} {\bibfnamefont {F.}~\bibnamefont {La~Via}},\ }\href {\doibase 10.1063/1.1500787} {\bibfield  {journal} {\bibinfo  {journal} {J. Appl. Phys.}\ }\textbf {\bibinfo {volume} {92}},\ \bibinfo {pages} {3147} (\bibinfo {year} {2002})}\BibitemShut {NoStop}%
\bibitem [{\citenamefont {Radermacher}\ \emph {et~al.}(1993)\citenamefont {Radermacher}, \citenamefont {Monroe}, \citenamefont {White}, \citenamefont {Short},\ and\ \citenamefont {Jebasinski}}]{Radermacher1993}%
  \BibitemOpen
  \bibfield  {author} {\bibinfo {author} {\bibfnamefont {K.}~\bibnamefont {Radermacher}}, \bibinfo {author} {\bibfnamefont {D.}~\bibnamefont {Monroe}}, \bibinfo {author} {\bibfnamefont {A.~E.}\ \bibnamefont {White}}, \bibinfo {author} {\bibfnamefont {K.~T.}\ \bibnamefont {Short}}, \ and\ \bibinfo {author} {\bibfnamefont {R.}~\bibnamefont {Jebasinski}},\ }\href {\doibase 10.1103/PhysRevB.48.8002} {\bibfield  {journal} {\bibinfo  {journal} {Phys. Rev. B}\ }\textbf {\bibinfo {volume} {48}},\ \bibinfo {pages} {8002} (\bibinfo {year} {1993})}\BibitemShut {NoStop}%
\bibitem [{\citenamefont {Mattheiss}\ and\ \citenamefont {Hamann}(1988)}]{Mattheiss1988}%
  \BibitemOpen
  \bibfield  {author} {\bibinfo {author} {\bibfnamefont {L.~F.}\ \bibnamefont {Mattheiss}}\ and\ \bibinfo {author} {\bibfnamefont {D.~R.}\ \bibnamefont {Hamann}},\ }\href {\doibase 10.1103/PhysRevB.37.10623} {\bibfield  {journal} {\bibinfo  {journal} {Phys. Rev. B}\ }\textbf {\bibinfo {volume} {37}},\ \bibinfo {pages} {10623} (\bibinfo {year} {1988})}\BibitemShut {NoStop}%
\bibitem [{\citenamefont {Kittel}(2005)}]{Kittel2005}%
  \BibitemOpen
  \bibfield  {author} {\bibinfo {author} {\bibfnamefont {C.}~\bibnamefont {Kittel}},\ }\href@noop {} {\emph {\bibinfo {title} {Introduction to solid state physics}}}\ (\bibinfo  {publisher} {John Wiley \& sons, inc},\ \bibinfo {year} {2005})\BibitemShut {NoStop}%
\bibitem [{\citenamefont {Zmuidzinas}(2012)}]{Zmuidzinas2012}%
  \BibitemOpen
  \bibfield  {author} {\bibinfo {author} {\bibfnamefont {J.}~\bibnamefont {Zmuidzinas}},\ }\href {\doibase 10.1146/annurev-conmatphys-020911-125022} {\bibfield  {journal} {\bibinfo  {journal} {Annu. Rev. Condens. Matter Phys.}\ }\textbf {\bibinfo {volume} {3}},\ \bibinfo {pages} {169} (\bibinfo {year} {2012})}\BibitemShut {NoStop}%
\bibitem [{\citenamefont {Newcombe}\ and\ \citenamefont {Lonzarich}(1988)}]{Newcombe1988}%
  \BibitemOpen
  \bibfield  {author} {\bibinfo {author} {\bibfnamefont {G.~C.~F.}\ \bibnamefont {Newcombe}}\ and\ \bibinfo {author} {\bibfnamefont {G.~G.}\ \bibnamefont {Lonzarich}},\ }\href {\doibase 10.1103/PhysRevB.37.10619} {\bibfield  {journal} {\bibinfo  {journal} {Phys. Rev. B}\ }\textbf {\bibinfo {volume} {37}},\ \bibinfo {pages} {10619} (\bibinfo {year} {1988})}\BibitemShut {NoStop}%
\bibitem [{\citenamefont {Lin}\ and\ \citenamefont {Bird}(2002)}]{JJLin2002}%
  \BibitemOpen
  \bibfield  {author} {\bibinfo {author} {\bibfnamefont {J.-J.}\ \bibnamefont {Lin}}\ and\ \bibinfo {author} {\bibfnamefont {J.}~\bibnamefont {Bird}},\ }\href {\doibase 10.1088/0953-8984/14/18/201} {\bibfield  {journal} {\bibinfo  {journal} {J. Phys. Condens. Matter.}\ }\textbf {\bibinfo {volume} {14}},\ \bibinfo {pages} {R501} (\bibinfo {year} {2002})}\BibitemShut {NoStop}%
\bibitem [{Note1()}]{Note1}%
  \BibitemOpen
  \bibinfo {note} {The relevant electronic parameters quoted in Ref. \cite {Chiu.21} were extracted using the mobility $\mu = e\tau _e/m^\ast $ dependence of the classical MR.}\BibitemShut {Stop}%
\bibitem [{\citenamefont {Sau}\ \emph {et~al.}(2010)\citenamefont {Sau}, \citenamefont {Tewari}, \citenamefont {Lutchyn}, \citenamefont {Stanescu},\ and\ \citenamefont {Sarma}}]{Sau2010}%
  \BibitemOpen
  \bibfield  {author} {\bibinfo {author} {\bibfnamefont {J.~D.}\ \bibnamefont {Sau}}, \bibinfo {author} {\bibfnamefont {S.}~\bibnamefont {Tewari}}, \bibinfo {author} {\bibfnamefont {R.~M.}\ \bibnamefont {Lutchyn}}, \bibinfo {author} {\bibfnamefont {T.~D.}\ \bibnamefont {Stanescu}}, \ and\ \bibinfo {author} {\bibfnamefont {S.~D.}\ \bibnamefont {Sarma}},\ }\href {\doibase 10.1103/PhysRevB.82.214509} {\bibfield  {journal} {\bibinfo  {journal} {Phys. Rev. B}\ }\textbf {\bibinfo {volume} {82}},\ \bibinfo {pages} {214509} (\bibinfo {year} {2010})}\BibitemShut {NoStop}%
\bibitem [{\citenamefont {Bergeret}\ and\ \citenamefont {Tokatly}(2014)}]{Tokatly2014}%
  \BibitemOpen
  \bibfield  {author} {\bibinfo {author} {\bibfnamefont {F.~S.}\ \bibnamefont {Bergeret}}\ and\ \bibinfo {author} {\bibfnamefont {I.~V.}\ \bibnamefont {Tokatly}},\ }\href {\doibase 10.1103/PhysRevB.89.134517} {\bibfield  {journal} {\bibinfo  {journal} {Phys. Rev. B}\ }\textbf {\bibinfo {volume} {89}},\ \bibinfo {pages} {134517} (\bibinfo {year} {2014})}\BibitemShut {NoStop}%
\bibitem [{\citenamefont {Manchon}\ \emph {et~al.}(2015)\citenamefont {Manchon}, \citenamefont {Koo}, \citenamefont {Nitta}, \citenamefont {Frolov},\ and\ \citenamefont {Duine}}]{Manchon2015}%
  \BibitemOpen
  \bibfield  {author} {\bibinfo {author} {\bibfnamefont {A.}~\bibnamefont {Manchon}}, \bibinfo {author} {\bibfnamefont {H.~C.}\ \bibnamefont {Koo}}, \bibinfo {author} {\bibfnamefont {J.}~\bibnamefont {Nitta}}, \bibinfo {author} {\bibfnamefont {S.~M.}\ \bibnamefont {Frolov}}, \ and\ \bibinfo {author} {\bibfnamefont {R.~A.}\ \bibnamefont {Duine}},\ }\href {\doibase 10.1038/nmat4360} {\bibfield  {journal} {\bibinfo  {journal} {Nat. Mater.}\ }\textbf {\bibinfo {volume} {14}},\ \bibinfo {pages} {871} (\bibinfo {year} {2015})}\BibitemShut {NoStop}%
\bibitem [{\citenamefont {Abrikosov}\ and\ \citenamefont {Gor’kov}(1962)}]{Abrikosov1962}%
  \BibitemOpen
  \bibfield  {author} {\bibinfo {author} {\bibfnamefont {A.~A.}\ \bibnamefont {Abrikosov}}\ and\ \bibinfo {author} {\bibfnamefont {L.~P.}\ \bibnamefont {Gor’kov}},\ }\href {http://jetp.ras.ru/cgi-bin/dn/e_015_04_0752.pdf} {\bibfield  {journal} {\bibinfo  {journal} {Sov. Phys. JETP}\ }\textbf {\bibinfo {volume} {15}},\ \bibinfo {pages} {752} (\bibinfo {year} {1962})}\BibitemShut {NoStop}%
\bibitem [{\citenamefont {Mishra}\ \emph {et~al.}(2021)\citenamefont {Mishra}, \citenamefont {Li}, \citenamefont {Zhang},\ and\ \citenamefont {Kirchner}}]{Mishra.21}%
  \BibitemOpen
  \bibfield  {author} {\bibinfo {author} {\bibfnamefont {V.}~\bibnamefont {Mishra}}, \bibinfo {author} {\bibfnamefont {Y.}~\bibnamefont {Li}}, \bibinfo {author} {\bibfnamefont {F.-C.}\ \bibnamefont {Zhang}}, \ and\ \bibinfo {author} {\bibfnamefont {S.}~\bibnamefont {Kirchner}},\ }\href {\doibase 10.1103/PhysRevB.103.184505} {\bibfield  {journal} {\bibinfo  {journal} {Phys. Rev. B}\ }\textbf {\bibinfo {volume} {103}},\ \bibinfo {pages} {184505} (\bibinfo {year} {2021})}\BibitemShut {NoStop}%
\bibitem [{\citenamefont {Hikami}\ \emph {et~al.}(1980)\citenamefont {Hikami}, \citenamefont {Larkin},\ and\ \citenamefont {Nagaoka}}]{Hikami1980}%
  \BibitemOpen
  \bibfield  {author} {\bibinfo {author} {\bibfnamefont {S.}~\bibnamefont {Hikami}}, \bibinfo {author} {\bibfnamefont {A.~I.}\ \bibnamefont {Larkin}}, \ and\ \bibinfo {author} {\bibfnamefont {Y.}~\bibnamefont {Nagaoka}},\ }\href {\doibase 10.1143/PTP.63.707} {\bibfield  {journal} {\bibinfo  {journal} {Prog. theor. phys.}\ }\textbf {\bibinfo {volume} {63}},\ \bibinfo {pages} {707} (\bibinfo {year} {1980})}\BibitemShut {NoStop}%
\bibitem [{\citenamefont {Wu}\ and\ \citenamefont {Lin}(1994)}]{Wu1994}%
  \BibitemOpen
  \bibfield  {author} {\bibinfo {author} {\bibfnamefont {C.~Y.}\ \bibnamefont {Wu}}\ and\ \bibinfo {author} {\bibfnamefont {J.~J.}\ \bibnamefont {Lin}},\ }\href {\doibase 10.1103/PhysRevB.50.385} {\bibfield  {journal} {\bibinfo  {journal} {Phys. Rev. B}\ }\textbf {\bibinfo {volume} {50}},\ \bibinfo {pages} {385} (\bibinfo {year} {1994})}\BibitemShut {NoStop}%
\bibitem [{\citenamefont {Fabian}\ and\ \citenamefont {Sarma}(1999)}]{Fabian1999}%
  \BibitemOpen
  \bibfield  {author} {\bibinfo {author} {\bibfnamefont {J.}~\bibnamefont {Fabian}}\ and\ \bibinfo {author} {\bibfnamefont {S.~D.}\ \bibnamefont {Sarma}},\ }\href {\doibase 10.1116/1.590813} {\bibfield  {journal} {\bibinfo  {journal} {J. Vac. Sci. Technol. B.}\ }\textbf {\bibinfo {volume} {17}},\ \bibinfo {pages} {1708} (\bibinfo {year} {1999})}\BibitemShut {NoStop}%
\bibitem [{\citenamefont {Santhanam}\ \emph {et~al.}(1987)\citenamefont {Santhanam}, \citenamefont {Wind},\ and\ \citenamefont {Prober}}]{Santhanam1987}%
  \BibitemOpen
  \bibfield  {author} {\bibinfo {author} {\bibfnamefont {P.}~\bibnamefont {Santhanam}}, \bibinfo {author} {\bibfnamefont {S.}~\bibnamefont {Wind}}, \ and\ \bibinfo {author} {\bibfnamefont {D.}~\bibnamefont {Prober}},\ }\href {\doibase 10.1103/PhysRevB.35.3188} {\bibfield  {journal} {\bibinfo  {journal} {Phys. Rev. B}\ }\textbf {\bibinfo {volume} {35}},\ \bibinfo {pages} {3188} (\bibinfo {year} {1987})}\BibitemShut {NoStop}%
\bibitem [{\citenamefont {Pierre}\ \emph {et~al.}(2003)\citenamefont {Pierre}, \citenamefont {Gougam}, \citenamefont {Anthore}, \citenamefont {Pothier}, \citenamefont {Esteve},\ and\ \citenamefont {Birge}}]{Pierre2003}%
  \BibitemOpen
  \bibfield  {author} {\bibinfo {author} {\bibfnamefont {F.}~\bibnamefont {Pierre}}, \bibinfo {author} {\bibfnamefont {A.~B.}\ \bibnamefont {Gougam}}, \bibinfo {author} {\bibfnamefont {A.}~\bibnamefont {Anthore}}, \bibinfo {author} {\bibfnamefont {H.}~\bibnamefont {Pothier}}, \bibinfo {author} {\bibfnamefont {D.}~\bibnamefont {Esteve}}, \ and\ \bibinfo {author} {\bibfnamefont {N.~O.}\ \bibnamefont {Birge}},\ }\href {\doibase 10.1103/PhysRevB.68.085413} {\bibfield  {journal} {\bibinfo  {journal} {Phys. Rev. B}\ }\textbf {\bibinfo {volume} {68}},\ \bibinfo {pages} {085413} (\bibinfo {year} {2003})}\BibitemShut {NoStop}%
\bibitem [{\citenamefont {Hsu}\ \emph {et~al.}(1999)\citenamefont {Hsu}, \citenamefont {Sheng},\ and\ \citenamefont {Lin}}]{Hsu1999}%
  \BibitemOpen
  \bibfield  {author} {\bibinfo {author} {\bibfnamefont {S.~Y.}\ \bibnamefont {Hsu}}, \bibinfo {author} {\bibfnamefont {P.~J.}\ \bibnamefont {Sheng}}, \ and\ \bibinfo {author} {\bibfnamefont {J.~J.}\ \bibnamefont {Lin}},\ }\href {\doibase 10.1103/PhysRevB.60.3940} {\bibfield  {journal} {\bibinfo  {journal} {Phys. Rev. B}\ }\textbf {\bibinfo {volume} {60}},\ \bibinfo {pages} {3940} (\bibinfo {year} {1999})}\BibitemShut {NoStop}%
\bibitem [{\citenamefont {Geier}\ and\ \citenamefont {Bergmann}(1992)}]{Bergmann1992}%
  \BibitemOpen
  \bibfield  {author} {\bibinfo {author} {\bibfnamefont {S.}~\bibnamefont {Geier}}\ and\ \bibinfo {author} {\bibfnamefont {G.}~\bibnamefont {Bergmann}},\ }\href {\doibase 10.1103/PhysRevLett.68.2520} {\bibfield  {journal} {\bibinfo  {journal} {Phys. Rev. Lett.}\ }\textbf {\bibinfo {volume} {68}},\ \bibinfo {pages} {2520} (\bibinfo {year} {1992})}\BibitemShut {NoStop}%
\bibitem [{\citenamefont {Mineev}\ and\ \citenamefont {Samokhin}(1994)}]{Mineev.94}%
  \BibitemOpen
  \bibfield  {author} {\bibinfo {author} {\bibfnamefont {V.~P.}\ \bibnamefont {Mineev}}\ and\ \bibinfo {author} {\bibfnamefont {K.~V.}\ \bibnamefont {Samokhin}},\ }\href {http://www.jetp.ras.ru/cgi-bin/dn/e_078_03_0401.pdf} {\bibfield  {journal} {\bibinfo  {journal} {Sov. Phys. JETP}\ }\textbf {\bibinfo {volume} {78}},\ \bibinfo {pages} {401} (\bibinfo {year} {1994})}\BibitemShut {NoStop}%
\bibitem [{\citenamefont {Annunziata}\ \emph {et~al.}(2012)\citenamefont {Annunziata}, \citenamefont {Manske},\ and\ \citenamefont {Linder}}]{Annunziata2012}%
  \BibitemOpen
  \bibfield  {author} {\bibinfo {author} {\bibfnamefont {G.}~\bibnamefont {Annunziata}}, \bibinfo {author} {\bibfnamefont {D.}~\bibnamefont {Manske}}, \ and\ \bibinfo {author} {\bibfnamefont {J.}~\bibnamefont {Linder}},\ }\href {\doibase 10.1103/PhysRevB.86.174514} {\bibfield  {journal} {\bibinfo  {journal} {Phys. Rev. B}\ }\textbf {\bibinfo {volume} {86}},\ \bibinfo {pages} {174514} (\bibinfo {year} {2012})}\BibitemShut {NoStop}%
\bibitem [{\citenamefont {DiTusa}\ \emph {et~al.}(1990)\citenamefont {DiTusa}, \citenamefont {Parpia},\ and\ \citenamefont {Phillips}}]{DiTusa1990}%
  \BibitemOpen
  \bibfield  {author} {\bibinfo {author} {\bibfnamefont {J.~F.}\ \bibnamefont {DiTusa}}, \bibinfo {author} {\bibfnamefont {J.~M.}\ \bibnamefont {Parpia}}, \ and\ \bibinfo {author} {\bibfnamefont {J.~M.}\ \bibnamefont {Phillips}},\ }\href {\doibase 10.1063/1.103663} {\bibfield  {journal} {\bibinfo  {journal} {Appl. Phys. Lett.}\ }\textbf {\bibinfo {volume} {57}},\ \bibinfo {pages} {452} (\bibinfo {year} {1990})}\BibitemShut {NoStop}%
\bibitem [{\citenamefont {Matsui}\ \emph {et~al.}(1990)\citenamefont {Matsui}, \citenamefont {Ohshima}, \citenamefont {Komori},\ and\ \citenamefont {Kobayashi}}]{Matsui1990}%
  \BibitemOpen
  \bibfield  {author} {\bibinfo {author} {\bibfnamefont {M.}~\bibnamefont {Matsui}}, \bibinfo {author} {\bibfnamefont {T.}~\bibnamefont {Ohshima}}, \bibinfo {author} {\bibfnamefont {F.}~\bibnamefont {Komori}}, \ and\ \bibinfo {author} {\bibfnamefont {S.}~\bibnamefont {Kobayashi}},\ }\href {\doibase 10.1063/1.345158} {\bibfield  {journal} {\bibinfo  {journal} {J. Appl. Phys.}\ }\textbf {\bibinfo {volume} {67}},\ \bibinfo {pages} {6368} (\bibinfo {year} {1990})}\BibitemShut {NoStop}%
\bibitem [{\citenamefont {Dutta}\ and\ \citenamefont {Horn}(1981)}]{Dutta1981}%
  \BibitemOpen
  \bibfield  {author} {\bibinfo {author} {\bibfnamefont {P.}~\bibnamefont {Dutta}}\ and\ \bibinfo {author} {\bibfnamefont {P.~M.}\ \bibnamefont {Horn}},\ }\href {\doibase 10.1103/RevModPhys.53.497} {\bibfield  {journal} {\bibinfo  {journal} {Rev. Mod. Phys.}\ }\textbf {\bibinfo {volume} {53}},\ \bibinfo {pages} {497} (\bibinfo {year} {1981})}\BibitemShut {NoStop}%
\bibitem [{\citenamefont {Yeh}\ \emph {et~al.}(2017)\citenamefont {Yeh}, \citenamefont {Chang},\ and\ \citenamefont {Lin}}]{Yeh2017}%
  \BibitemOpen
  \bibfield  {author} {\bibinfo {author} {\bibfnamefont {S.-S.}\ \bibnamefont {Yeh}}, \bibinfo {author} {\bibfnamefont {W.-Y.}\ \bibnamefont {Chang}}, \ and\ \bibinfo {author} {\bibfnamefont {J.-J.}\ \bibnamefont {Lin}},\ }\href {\doibase 10.1126/sciadv.1700135} {\bibfield  {journal} {\bibinfo  {journal} {Sci. Adv.}\ }\textbf {\bibinfo {volume} {3}},\ \bibinfo {pages} {e1700135} (\bibinfo {year} {2017})}\BibitemShut {NoStop}%
\bibitem [{\citenamefont {Yeh}\ \emph {et~al.}(2018)\citenamefont {Yeh}, \citenamefont {Gao}, \citenamefont {Wu}, \citenamefont {Su},\ and\ \citenamefont {Lin}}]{Yeh2018}%
  \BibitemOpen
  \bibfield  {author} {\bibinfo {author} {\bibfnamefont {S.-S.}\ \bibnamefont {Yeh}}, \bibinfo {author} {\bibfnamefont {K.~H.}\ \bibnamefont {Gao}}, \bibinfo {author} {\bibfnamefont {T.-L.}\ \bibnamefont {Wu}}, \bibinfo {author} {\bibfnamefont {T.-K.}\ \bibnamefont {Su}}, \ and\ \bibinfo {author} {\bibfnamefont {J.-J.}\ \bibnamefont {Lin}},\ }\href {\doibase 10.1103/PhysRevApplied.10.034004} {\bibfield  {journal} {\bibinfo  {journal} {Phys. Rev. Appl.}\ }\textbf {\bibinfo {volume} {10}},\ \bibinfo {pages} {034004} (\bibinfo {year} {2018})}\BibitemShut {NoStop}%
\bibitem [{\citenamefont {van~den Homberg}\ \emph {et~al.}(1998)\citenamefont {van~den Homberg}, \citenamefont {Verbruggen}, \citenamefont {Alkemade}, \citenamefont {Radelaar}, \citenamefont {Ochs}, \citenamefont {Armbruster-Dagge}, \citenamefont {Seeger},\ and\ \citenamefont {Stoll}}]{Van1998}%
  \BibitemOpen
  \bibfield  {author} {\bibinfo {author} {\bibfnamefont {M.~J.~C.}\ \bibnamefont {van~den Homberg}}, \bibinfo {author} {\bibfnamefont {A.~H.}\ \bibnamefont {Verbruggen}}, \bibinfo {author} {\bibfnamefont {P.~F.~A.}\ \bibnamefont {Alkemade}}, \bibinfo {author} {\bibfnamefont {S.}~\bibnamefont {Radelaar}}, \bibinfo {author} {\bibfnamefont {E.}~\bibnamefont {Ochs}}, \bibinfo {author} {\bibfnamefont {K.}~\bibnamefont {Armbruster-Dagge}}, \bibinfo {author} {\bibfnamefont {A.}~\bibnamefont {Seeger}}, \ and\ \bibinfo {author} {\bibfnamefont {H.}~\bibnamefont {Stoll}},\ }\href {\doibase 10.1103/PhysRevB.57.53} {\bibfield  {journal} {\bibinfo  {journal} {Phys. Rev. B}\ }\textbf {\bibinfo {volume} {57}},\ \bibinfo {pages} {53} (\bibinfo {year} {1998})}\BibitemShut {NoStop}%
\bibitem [{\citenamefont {Scofield}\ \emph {et~al.}(1985)\citenamefont {Scofield}, \citenamefont {Mantese},\ and\ \citenamefont {Webb}}]{Scofield1985}%
  \BibitemOpen
  \bibfield  {author} {\bibinfo {author} {\bibfnamefont {J.~H.}\ \bibnamefont {Scofield}}, \bibinfo {author} {\bibfnamefont {J.~V.}\ \bibnamefont {Mantese}}, \ and\ \bibinfo {author} {\bibfnamefont {W.~W.}\ \bibnamefont {Webb}},\ }\href {\doibase 10.1103/PhysRevB.32.736} {\bibfield  {journal} {\bibinfo  {journal} {Phys. Rev. B}\ }\textbf {\bibinfo {volume} {32}},\ \bibinfo {pages} {736} (\bibinfo {year} {1985})}\BibitemShut {NoStop}%
\bibitem [{\citenamefont {Mukhanova}\ \emph {et~al.}(2023)\citenamefont {Mukhanova}, \citenamefont {Zeng}, \citenamefont {Heredia}, \citenamefont {Wu}, \citenamefont {Lilja}, \citenamefont {Lin}, \citenamefont {Yeh},\ and\ \citenamefont {Hakonen}}]{Mukhanova2023}%
  \BibitemOpen
  \bibfield  {author} {\bibinfo {author} {\bibfnamefont {E.}~\bibnamefont {Mukhanova}}, \bibinfo {author} {\bibfnamefont {W.}~\bibnamefont {Zeng}}, \bibinfo {author} {\bibfnamefont {E.~A.}\ \bibnamefont {Heredia}}, \bibinfo {author} {\bibfnamefont {C.-W.}\ \bibnamefont {Wu}}, \bibinfo {author} {\bibfnamefont {I.}~\bibnamefont {Lilja}}, \bibinfo {author} {\bibfnamefont {J.-J.}\ \bibnamefont {Lin}}, \bibinfo {author} {\bibfnamefont {S.-S.}\ \bibnamefont {Yeh}}, \ and\ \bibinfo {author} {\bibfnamefont {P.}~\bibnamefont {Hakonen}},\ }\href@noop {} {} (\bibinfo {year} {2023}),\ \Eprint {http://arxiv.org/abs/2312.14624} {arXiv:2312.14624 [cond-mat.supr-con]} \BibitemShut {NoStop}%
\bibitem [{\citenamefont {Cohn}\ \emph {et~al.}(1988)\citenamefont {Cohn}, \citenamefont {Lin}, \citenamefont {Lamelas}, \citenamefont {He}, \citenamefont {Clarke},\ and\ \citenamefont {Uher}}]{Cohn1988}%
  \BibitemOpen
  \bibfield  {author} {\bibinfo {author} {\bibfnamefont {J.~L.}\ \bibnamefont {Cohn}}, \bibinfo {author} {\bibfnamefont {J.~J.}\ \bibnamefont {Lin}}, \bibinfo {author} {\bibfnamefont {F.~J.}\ \bibnamefont {Lamelas}}, \bibinfo {author} {\bibfnamefont {H.}~\bibnamefont {He}}, \bibinfo {author} {\bibfnamefont {R.}~\bibnamefont {Clarke}}, \ and\ \bibinfo {author} {\bibfnamefont {C.}~\bibnamefont {Uher}},\ }\href {\doibase 10.1103/PhysRevB.38.2326} {\bibfield  {journal} {\bibinfo  {journal} {Phys. Rev. B}\ }\textbf {\bibinfo {volume} {38}},\ \bibinfo {pages} {2326} (\bibinfo {year} {1988})}\BibitemShut {NoStop}%
\bibitem [{\citenamefont {Tinkham}(2004)}]{Tinkham2004}%
  \BibitemOpen
  \bibfield  {author} {\bibinfo {author} {\bibfnamefont {M.}~\bibnamefont {Tinkham}},\ }\href@noop {} {\emph {\bibinfo {title} {Introduction to superconductivity}}}\ (\bibinfo  {publisher} {Courier Corporation},\ \bibinfo {year} {2004})\BibitemShut {NoStop}%
\bibitem [{\citenamefont {Gubin}\ \emph {et~al.}(2005)\citenamefont {Gubin}, \citenamefont {Il’in}, \citenamefont {Vitusevich}, \citenamefont {Siegel},\ and\ \citenamefont {Klein}}]{Gubin2005}%
  \BibitemOpen
  \bibfield  {author} {\bibinfo {author} {\bibfnamefont {A.~I.}\ \bibnamefont {Gubin}}, \bibinfo {author} {\bibfnamefont {K.~S.}\ \bibnamefont {Il’in}}, \bibinfo {author} {\bibfnamefont {S.~A.}\ \bibnamefont {Vitusevich}}, \bibinfo {author} {\bibfnamefont {M.}~\bibnamefont {Siegel}}, \ and\ \bibinfo {author} {\bibfnamefont {N.}~\bibnamefont {Klein}},\ }\href {\doibase 10.1103/PhysRevB.72.064503} {\bibfield  {journal} {\bibinfo  {journal} {Phys. Rev. B}\ }\textbf {\bibinfo {volume} {72}},\ \bibinfo {pages} {064503} (\bibinfo {year} {2005})}\BibitemShut {NoStop}%
\bibitem [{\citenamefont {Fuchs}\ \emph {et~al.}(2022)\citenamefont {Fuchs}, \citenamefont {Kochan}, \citenamefont {Schmidt}, \citenamefont {H\"uttner}, \citenamefont {Baumgartner}, \citenamefont {Reinhardt}, \citenamefont {Gronin}, \citenamefont {Gardner}, \citenamefont {Lindemann}, \citenamefont {Manfra}, \citenamefont {Strunk},\ and\ \citenamefont {Paradiso}}]{Fuchs2022}%
  \BibitemOpen
  \bibfield  {author} {\bibinfo {author} {\bibfnamefont {L.}~\bibnamefont {Fuchs}}, \bibinfo {author} {\bibfnamefont {D.}~\bibnamefont {Kochan}}, \bibinfo {author} {\bibfnamefont {J.}~\bibnamefont {Schmidt}}, \bibinfo {author} {\bibfnamefont {N.}~\bibnamefont {H\"uttner}}, \bibinfo {author} {\bibfnamefont {C.}~\bibnamefont {Baumgartner}}, \bibinfo {author} {\bibfnamefont {S.}~\bibnamefont {Reinhardt}}, \bibinfo {author} {\bibfnamefont {S.}~\bibnamefont {Gronin}}, \bibinfo {author} {\bibfnamefont {G.~C.}\ \bibnamefont {Gardner}}, \bibinfo {author} {\bibfnamefont {T.}~\bibnamefont {Lindemann}}, \bibinfo {author} {\bibfnamefont {M.~J.}\ \bibnamefont {Manfra}}, \bibinfo {author} {\bibfnamefont {C.}~\bibnamefont {Strunk}}, \ and\ \bibinfo {author} {\bibfnamefont {N.}~\bibnamefont {Paradiso}},\ }\href {\doibase 10.1103/PhysRevX.12.041020} {\bibfield  {journal} {\bibinfo  {journal} {Phys. Rev. X}\ }\textbf {\bibinfo {volume} {12}},\ \bibinfo {pages} {041020} (\bibinfo {year} {2022})}\BibitemShut {NoStop}%
\bibitem [{\citenamefont {Scharnberg}\ and\ \citenamefont {Klemm}(1980)}]{Scharnberg.80}%
  \BibitemOpen
  \bibfield  {author} {\bibinfo {author} {\bibfnamefont {K.}~\bibnamefont {Scharnberg}}\ and\ \bibinfo {author} {\bibfnamefont {R.~A.}\ \bibnamefont {Klemm}},\ }\href {\doibase 10.1103/PhysRevB.22.5233} {\bibfield  {journal} {\bibinfo  {journal} {Phys. Rev. B}\ }\textbf {\bibinfo {volume} {22}},\ \bibinfo {pages} {5233} (\bibinfo {year} {1980})}\BibitemShut {NoStop}%
\bibitem [{\citenamefont {Burlachkov}(1985)}]{Burlachkov.85}%
  \BibitemOpen
  \bibfield  {author} {\bibinfo {author} {\bibfnamefont {L.~I.}\ \bibnamefont {Burlachkov}},\ }\href {http://jetp.ras.ru/cgi-bin/dn/e_062_04_0800.pdf} {\bibfield  {journal} {\bibinfo  {journal} {Sov. Phys. JETP}\ }\textbf {\bibinfo {volume} {62}},\ \bibinfo {pages} {800} (\bibinfo {year} {1985})}\BibitemShut {NoStop}%
\bibitem [{\citenamefont {Mishra}\ \emph {et~al.}(2023)\citenamefont {Mishra}, \citenamefont {Li}, \citenamefont {Zhang},\ and\ \citenamefont {Kirchner}}]{Mishra.23}%
  \BibitemOpen
  \bibfield  {author} {\bibinfo {author} {\bibfnamefont {V.}~\bibnamefont {Mishra}}, \bibinfo {author} {\bibfnamefont {Y.}~\bibnamefont {Li}}, \bibinfo {author} {\bibfnamefont {F.-C.}\ \bibnamefont {Zhang}}, \ and\ \bibinfo {author} {\bibfnamefont {S.}~\bibnamefont {Kirchner}},\ }\href {\doibase 10.1103/PhysRevB.107.184505} {\bibfield  {journal} {\bibinfo  {journal} {Phys. Rev. B}\ }\textbf {\bibinfo {volume} {107}},\ \bibinfo {pages} {184505} (\bibinfo {year} {2023})}\BibitemShut {NoStop}%
\bibitem [{\citenamefont {Ruvalds}(1996)}]{Ruvalds1996}%
  \BibitemOpen
  \bibfield  {author} {\bibinfo {author} {\bibfnamefont {J.}~\bibnamefont {Ruvalds}},\ }\href {\doibase 10.1088/0953-2048/9/11/001} {\bibfield  {journal} {\bibinfo  {journal} {Supercond Sci Technol.}\ }\textbf {\bibinfo {volume} {9}},\ \bibinfo {pages} {905} (\bibinfo {year} {1996})}\BibitemShut {NoStop}%
\bibitem [{\citenamefont {Wallraff}\ \emph {et~al.}(2004)\citenamefont {Wallraff}, \citenamefont {Schuster}, \citenamefont {Blais}, \citenamefont {Frunzio}, \citenamefont {Huang}, \citenamefont {Majer}, \citenamefont {Kumar}, \citenamefont {Girvin},\ and\ \citenamefont {Schoelkopf}}]{Wallraff2004}%
  \BibitemOpen
  \bibfield  {author} {\bibinfo {author} {\bibfnamefont {A.}~\bibnamefont {Wallraff}}, \bibinfo {author} {\bibfnamefont {D.~I.}\ \bibnamefont {Schuster}}, \bibinfo {author} {\bibfnamefont {A.}~\bibnamefont {Blais}}, \bibinfo {author} {\bibfnamefont {L.}~\bibnamefont {Frunzio}}, \bibinfo {author} {\bibfnamefont {R.-S.}\ \bibnamefont {Huang}}, \bibinfo {author} {\bibfnamefont {J.}~\bibnamefont {Majer}}, \bibinfo {author} {\bibfnamefont {S.}~\bibnamefont {Kumar}}, \bibinfo {author} {\bibfnamefont {S.~M.}\ \bibnamefont {Girvin}}, \ and\ \bibinfo {author} {\bibfnamefont {R.~J.}\ \bibnamefont {Schoelkopf}},\ }\href {\doibase https://doi.org/10.1038/nature02851} {\bibfield  {journal} {\bibinfo  {journal} {Nature}\ }\textbf {\bibinfo {volume} {431}},\ \bibinfo {pages} {162} (\bibinfo {year} {2004})}\BibitemShut {NoStop}%
\bibitem [{\citenamefont {Mariantoni}\ \emph {et~al.}(2011)\citenamefont {Mariantoni}, \citenamefont {Wang}, \citenamefont {Yamamoto}, \citenamefont {Neeley}, \citenamefont {Bialczak}, \citenamefont {Chen}, \citenamefont {Lenander}, \citenamefont {Lucero}, \citenamefont {O’Connell}, \citenamefont {Sank}, \citenamefont {Weides}, \citenamefont {Wenner}, \citenamefont {Yin}, \citenamefont {Zhao}, \citenamefont {Korotkov}, \citenamefont {Cleland},\ and\ \citenamefont {Martinis}}]{Mariantoni2011}%
  \BibitemOpen
  \bibfield  {author} {\bibinfo {author} {\bibfnamefont {M.}~\bibnamefont {Mariantoni}}, \bibinfo {author} {\bibfnamefont {H.}~\bibnamefont {Wang}}, \bibinfo {author} {\bibfnamefont {T.}~\bibnamefont {Yamamoto}}, \bibinfo {author} {\bibfnamefont {M.}~\bibnamefont {Neeley}}, \bibinfo {author} {\bibfnamefont {R.~C.}\ \bibnamefont {Bialczak}}, \bibinfo {author} {\bibfnamefont {Y.}~\bibnamefont {Chen}}, \bibinfo {author} {\bibfnamefont {M.}~\bibnamefont {Lenander}}, \bibinfo {author} {\bibfnamefont {E.}~\bibnamefont {Lucero}}, \bibinfo {author} {\bibfnamefont {A.~D.}\ \bibnamefont {O’Connell}}, \bibinfo {author} {\bibfnamefont {D.}~\bibnamefont {Sank}}, \bibinfo {author} {\bibfnamefont {M.}~\bibnamefont {Weides}}, \bibinfo {author} {\bibfnamefont {J.}~\bibnamefont {Wenner}}, \bibinfo {author} {\bibfnamefont {Y.}~\bibnamefont {Yin}}, \bibinfo {author} {\bibfnamefont {J.}~\bibnamefont {Zhao}}, \bibinfo {author} {\bibfnamefont {A.~N.}\ \bibnamefont {Korotkov}}, \bibinfo {author} {\bibfnamefont {A.~N.}\
  \bibnamefont {Cleland}}, \ and\ \bibinfo {author} {\bibfnamefont {J.~M.}\ \bibnamefont {Martinis}},\ }\href {\doibase 10.1126/science.1208517} {\bibfield  {journal} {\bibinfo  {journal} {Science}\ }\textbf {\bibinfo {volume} {334}},\ \bibinfo {pages} {61} (\bibinfo {year} {2011})}\BibitemShut {NoStop}%
\bibitem [{\citenamefont {Hofheinz}\ \emph {et~al.}(2009)\citenamefont {Hofheinz}, \citenamefont {Wang}, \citenamefont {Ansmann}, \citenamefont {Bialczak}, \citenamefont {Lucero}, \citenamefont {Neeley}, \citenamefont {O'connell}, \citenamefont {Sank}, \citenamefont {Wenner}, \citenamefont {Martinis},\ and\ \citenamefont {Cleland}}]{Hofheinz2009}%
  \BibitemOpen
  \bibfield  {author} {\bibinfo {author} {\bibfnamefont {M.}~\bibnamefont {Hofheinz}}, \bibinfo {author} {\bibfnamefont {H.}~\bibnamefont {Wang}}, \bibinfo {author} {\bibfnamefont {M.}~\bibnamefont {Ansmann}}, \bibinfo {author} {\bibfnamefont {R.~C.}\ \bibnamefont {Bialczak}}, \bibinfo {author} {\bibfnamefont {E.}~\bibnamefont {Lucero}}, \bibinfo {author} {\bibfnamefont {M.}~\bibnamefont {Neeley}}, \bibinfo {author} {\bibfnamefont {A.~D.}\ \bibnamefont {O'connell}}, \bibinfo {author} {\bibfnamefont {D.}~\bibnamefont {Sank}}, \bibinfo {author} {\bibfnamefont {J.}~\bibnamefont {Wenner}}, \bibinfo {author} {\bibfnamefont {J.~M.}\ \bibnamefont {Martinis}}, \ and\ \bibinfo {author} {\bibfnamefont {A.~N.}\ \bibnamefont {Cleland}},\ }\href {\doibase https://doi.org/10.1038/nature08005} {\bibfield  {journal} {\bibinfo  {journal} {Nature}\ }\textbf {\bibinfo {volume} {459}},\ \bibinfo {pages} {546} (\bibinfo {year} {2009})}\BibitemShut {NoStop}%
\bibitem [{\citenamefont {Day}\ \emph {et~al.}(2003)\citenamefont {Day}, \citenamefont {LeDuc}, \citenamefont {Mazin}, \citenamefont {Vayonakis},\ and\ \citenamefont {Zmuidzinas}}]{Day2003}%
  \BibitemOpen
  \bibfield  {author} {\bibinfo {author} {\bibfnamefont {P.~K.}\ \bibnamefont {Day}}, \bibinfo {author} {\bibfnamefont {H.~G.}\ \bibnamefont {LeDuc}}, \bibinfo {author} {\bibfnamefont {B.~A.}\ \bibnamefont {Mazin}}, \bibinfo {author} {\bibfnamefont {A.}~\bibnamefont {Vayonakis}}, \ and\ \bibinfo {author} {\bibfnamefont {J.}~\bibnamefont {Zmuidzinas}},\ }\href {\doibase https://doi.org/10.1038/nature02037} {\bibfield  {journal} {\bibinfo  {journal} {Nature}\ }\textbf {\bibinfo {volume} {425}},\ \bibinfo {pages} {817} (\bibinfo {year} {2003})}\BibitemShut {NoStop}%
\bibitem [{\citenamefont {McRae}\ \emph {et~al.}(2020)\citenamefont {McRae}, \citenamefont {Wang}, \citenamefont {Gao}, \citenamefont {Vissers}, \citenamefont {Brecht}, \citenamefont {Dunsworth}, \citenamefont {Pappas},\ and\ \citenamefont {Mutus}}]{McRae2020}%
  \BibitemOpen
  \bibfield  {author} {\bibinfo {author} {\bibfnamefont {C.~R.~H.}\ \bibnamefont {McRae}}, \bibinfo {author} {\bibfnamefont {H.}~\bibnamefont {Wang}}, \bibinfo {author} {\bibfnamefont {J.}~\bibnamefont {Gao}}, \bibinfo {author} {\bibfnamefont {M.~R.}\ \bibnamefont {Vissers}}, \bibinfo {author} {\bibfnamefont {T.}~\bibnamefont {Brecht}}, \bibinfo {author} {\bibfnamefont {A.}~\bibnamefont {Dunsworth}}, \bibinfo {author} {\bibfnamefont {D.~P.}\ \bibnamefont {Pappas}}, \ and\ \bibinfo {author} {\bibfnamefont {J.}~\bibnamefont {Mutus}},\ }\href {\doibase 10.1063/5.0017378} {\bibfield  {journal} {\bibinfo  {journal} {Rev. Sci. Instrum.}\ }\textbf {\bibinfo {volume} {91}},\ \bibinfo {pages} {091101} (\bibinfo {year} {2020})}\BibitemShut {NoStop}%
\bibitem [{\citenamefont {Stern}\ and\ \citenamefont {Lindner}(2013)}]{Stern2013}%
  \BibitemOpen
  \bibfield  {author} {\bibinfo {author} {\bibfnamefont {A.}~\bibnamefont {Stern}}\ and\ \bibinfo {author} {\bibfnamefont {N.~H.}\ \bibnamefont {Lindner}},\ }\href@noop {} {\bibfield  {journal} {\bibinfo  {journal} {Science}\ }\textbf {\bibinfo {volume} {339}},\ \bibinfo {pages} {1179} (\bibinfo {year} {2013})}\BibitemShut {NoStop}%
\bibitem [{\citenamefont {Tang}\ \emph {et~al.}(2017)\citenamefont {Tang}, \citenamefont {Zhou},\ and\ \citenamefont {Zhang}}]{Tang2017}%
  \BibitemOpen
  \bibfield  {author} {\bibinfo {author} {\bibfnamefont {P.}~\bibnamefont {Tang}}, \bibinfo {author} {\bibfnamefont {Q.}~\bibnamefont {Zhou}}, \ and\ \bibinfo {author} {\bibfnamefont {S.-C.}\ \bibnamefont {Zhang}},\ }\href {\doibase 10.1103/PhysRevLett.119.206402} {\bibfield  {journal} {\bibinfo  {journal} {Phys. Rev. Lett.}\ }\textbf {\bibinfo {volume} {119}},\ \bibinfo {pages} {206402} (\bibinfo {year} {2017})}\BibitemShut {NoStop}%
\bibitem [{\citenamefont {Pshenay-Severin}\ \emph {et~al.}(2018)\citenamefont {Pshenay-Severin}, \citenamefont {Ivanov}, \citenamefont {Burkov},\ and\ \citenamefont {Burkov}}]{Pshenay2018}%
  \BibitemOpen
  \bibfield  {author} {\bibinfo {author} {\bibfnamefont {D.~A.}\ \bibnamefont {Pshenay-Severin}}, \bibinfo {author} {\bibfnamefont {Y.~V.}\ \bibnamefont {Ivanov}}, \bibinfo {author} {\bibfnamefont {A.~A.}\ \bibnamefont {Burkov}}, \ and\ \bibinfo {author} {\bibfnamefont {A.~T.}\ \bibnamefont {Burkov}},\ }\href {\doibase 10.1088/1361-648X/aab0ba} {\bibfield  {journal} {\bibinfo  {journal} {J. Phys. Condens. Matter.}\ }\textbf {\bibinfo {volume} {30}},\ \bibinfo {pages} {135501} (\bibinfo {year} {2018})}\BibitemShut {NoStop}%

\end{thebibliography}
\end{document}